\documentclass[11pt,english,fleqn]{article}
\usepackage[T1]{fontenc}
\usepackage{authblk}
\usepackage[utf8]{inputenc}
\usepackage[english]{babel}
\usepackage{graphicx}
\usepackage[round, sectionbib,authoryear]{natbib}     
\usepackage[labelformat=simple]{subcaption}
\usepackage{tabularx}
\usepackage{multirow}
\usepackage{graphicx}
\usepackage{psfrag}
\usepackage{amsmath}
\usepackage{amssymb}
\usepackage{nicefrac}
\usepackage[]{algorithm2e}
\usepackage{fullpage}
\usepackage{listings}
\usepackage{enumerate}
\usepackage{wrapfig}
\usepackage{float}
\usepackage{parskip}
\usepackage{lipsum}
\usepackage{listings,color}
\usepackage[font=small,labelfont=normal,labelsep=default]{caption}
\usepackage[nodisplayskipstretch]{setspace}
\usepackage{epstopdf}
\AppendGraphicsExtensions{.tif}
\usepackage[colorlinks=true,urlcolor=blue,citecolor=red,linkcolor=red,bookmarks=true]{hyperref} 
\usepackage{cleveref}
\usepackage{pstricks,pstricks-add,pst-3d,pstricks-add,pst-grad,multido}

\definecolor{halfgray}{gray}{0.55} 
\definecolor{webgreen}{rgb}{0,.5,0}
\definecolor{webbrown}{rgb}{.6,0,0}
\definecolor{BlueLUH}{cmyk}{1.0,0.7,0,0}
\colorlet{LightBlue}{BlueLUH!20!white}
\colorlet{DarkBlue}{BlueLUH!80!black!20}

\title{Dynamics of Argon Gas Bubbles Rising in Liquid Steel in the Presence of Transverse Magnetic Field}

\author[1]{Purushotam Kumar\thanks{pkumar8@illinois.edu}}
\affil[1]{Corning Incorporated, Manufacturing Technology \& Engineering, Corning, NY 14830}
\author[2]{Surya P. Vanka\thanks{spvanka@illinois.edu}}
\affil[2]{Mechanical Science \& Engineering, University of Illinois Urbana-Champaign, Urbana, IL 61801}

\date{\today}
\providecommand{\keywords}[1]{Keywords: #1}

\begin{document}

\maketitle
\onehalfspacing

\begin{abstract}	
	{\it Bubbly flows are present in various industrial processes including metallurgical processes in which gas bubbles are injected at the bottom of bulk liquid metal to stir the liquid metal and homogenize the metal. Understanding the motion of such bubbles is essential, as it has been shown that bubble flotation can remove inclusions. In this work, we have numerically studied three-dimensional dynamics of a pair of inline Argon bubbles rising in molten steel under the influence of a transverse magnetic field. We have explored the effects of two transverse magnetic field strengths (Bx = 0 and 0.2 T). The bubbles' motion and transient rise velocities are compared under different magnetic fields. The shape deformations and path of the bubbles are discussed. The flow structures behind the bubbles are analyzed. We found that structures are more organized and elongated under a magnetic field, whereas it is complex and intertwined when the magnetic field is not included. We have used a geometry construction-based volume of fluid (VOF) method to track interface, maintain mass balance and estimate the interface curvature. Additionally, we have incorporated a Sharp Surface Force Method (SSF) for surface tension forces. The algorithm is able to minimize the spurious velocities.
		
		
}  
\end{abstract}


\section{INTRODUCTION}
\label{sec:introduction}
Bubbly flows are encountered in various industrial processes and everyday life. To mix and homogenize the metal in metallurgical processes, gas bubbles are injected at the bottom of bulk liquid metal to stir the liquid metal. In the process for continuous casting of steel, which is widely used for steel making, Argon bubbles are commonly injected during the casting process. Understanding the motion of such argon bubbles is essential as it has been shown that bubble flotation can remove inclusions. In addition, in order to improve the product quality frequently an external magnetic field is applied to control the fluid motion and bubble behavior. In the past several decades, numerous theoretical, experimental, and computational studies have been carried out on the dynamics of a rising bubble in transparent liquids (such as water and oils). However, only a few studies have been reported on bubble motion in liquids when subjected to an external magnetic field. \par

Dynamics of a rising bubble are a function of the E\"otv\"os number ($Eo = \Delta \rho g d^2 \sigma^{-1}$, also referred as Bond number $Bo = \rho_l g d^2 \sigma^{-1}$), Morton number $(Mo = g \mu_o^4 \rho_l^{-1} \sigma^{-3})$, confinement ratio $(C_r = Wd^{-1})$, Hartmann number $Ha = \textbf{B}d\sqrt{\sigma/\mu_l}$ \citep{Jin2016mhd_bubble,Kumar2015non_newtonian}. An exhaustive collection of experimental data for a bubble rising in an unconfined medium has been summarized in works of \citet{Grace1973,Grace1976,Bhaga1981}, available as nomogram charts of the terminal Reynolds and bubble shape for combinations of Bond and Morton numbers. In addition, the proximity of the bubble to a wall alters the bubble rise velocity and shape and increases the shear force at the gas-liquid interface \cite{Figueroa-Espinoza2008,Kumar2015confinement}. Detailed reviews of bubble dynamics in Newtonian fluids are available in the works of \citet{Clift1978,DeKee2002}.

The dynamics of multiple bubbles rising in Newtonian fluids without magnetic field has been studied by several authors \citep{Abbassi2018,Gumulya2017,Watanabe2006,yuan_prosperetti_1994,legendre_magnaudet_mougin_2003,Yu2011LBM,KATZ1996239,RUZICKA20001141} and have reported that flow structure around the bubble has an important impact on the interaction between leading and trailing bubbles. They have reported that wake in front of a trailing bubble plays an important role in reducing the drag around it. Due to reduced drag, the trailing bubble accelerates until it coalesces with the leading bubble. 

%
%
%
In the present paper, we have briefly discussed a numerical technique to simulate two-phase flows and used it to study the three-dimensional deformation and rise of two equally sized bubbles in molten steel under the influence of a transverse magnetic field for a given bubble size and center to center distance. Various quantities such as the bubble shape, rise velocity, and rise paths are investigated. The algorithm used for the computations is presented in the numerical method section. The problem description is provided in the computational details section. In the results and discussion section, we present the results of the simulations and discuss the critical findings. A summary of the present findings is given at the end. \par

\section{Problem Setup}
\label{sec:computational_details}

\begin{figure}[h]
	\begin{center}
		\includegraphics[height=0.4\textwidth]{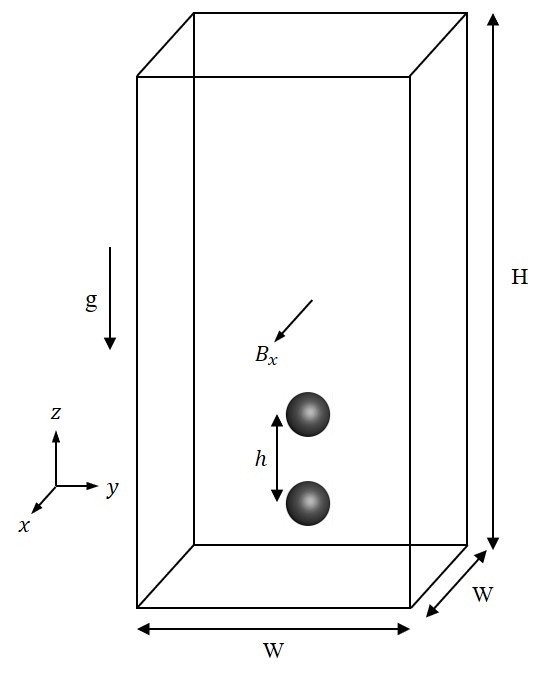}
	\end{center}
	\caption{Initial position of bubbles in liquid column}
	\label{fig:problem_statement}
\end{figure}

We consider a pair of deformable Argon bubbles rising in a rectangular column filled with liquid steel. The initial size $(d)$ and shape of both bubbles are identical and spherical, and the initial distance between their centers $(h)$ is adjusted. The trailing bubble is initially placed two diameters above the bottom boundary at the center of duct cross-section, and the leading bubble is placed at center to center distance of $h$. Both fluids (liquid in the column and bubbles) are stationary at the beginning of solution procedure. A constant transverse magnetic field of a given strength $(B_x)$ is applied in the $x-$ direction. The gravitational force acts downward along the negative $z-$axis. We have considered a three-dimensional domain of $4d \times 4d \times 24d$ along $x$, $y$ and $z$ directions with all boundaries of the domain to be no-slip, no-penetration and non-conductive walls. Based on a systematic grid refinement study, we have selected $32$ control volumes per bubble diameter as an adequate resolution, with a grid of $128 \times 128 \times 768$ ($\approx$ 13 million) control volumes for current study. Figure \ref{fig:problem_statement} shows sketch of the computational domain. 

Material properties for steel and Argon are taken as density: 7000 and 0.56 $kg/m^3$, viscosity: $6.3 \times 10 ^{-3}$ and $7.42 \times 10^{-5} Pa~s$, and electrical conductivity: 714000 and $10^{-15}$ $1/(\Omega ~s)$. The surface tension $(\gamma)$ is prescribed a value of $1.2 ~N/m$. \par

\section{Numerical method}

\subsection*{Governing equations}
We have used continuity, momentum and electromagnetic equations given by eq.\eqref{eqn:continuity}, \eqref{eqn:momentum_equation}, and \eqref{eqn:emf_ohms}-\eqref{eqn:emf_phi}, to simulate bubble dynamics. We assumed the flow to be incompressible and isothermal (ignore any heat generation due to viscous dissipation or eddy currents).  
\begin{alignat}{1}
	\label{eqn:continuity}
	\nabla \cdot  \mathbf{u} = 0
\end{alignat}
\vspace{-8mm}
\begin{equation}
	\label{eqn:momentum_equation}
	\begin{aligned}
		\frac{\partial \left(\rho \mathbf{u} \right)}{\partial t} + \nabla \cdot \left( \rho \mathbf{uu} \right) =  -\nabla p + \nabla \cdot \left( \mu \left[ \nabla \mathbf{u} + \nabla \mathbf{u}^{T} \right] \right) + \rho \textbf{g} + \gamma \kappa \mathbf{n} \delta \left( \mathbf{x} - \mathbf{x}_f \right) + \textbf{F}_L
	\end{aligned}
\end{equation}
\vspace{-8mm}
%
%
\begin{alignat}{2}
	\label{eqn:emf_ohms}
	\textbf{J} &= \sigma \left( -\nabla\Phi + \textbf{u}\times \textbf{B} \right) \\ 
	\label{eqn:emf_phi}
	\nabla \cdot \left( \sigma \nabla\Phi \right) &= \nabla \cdot \left[\sigma \left(\textbf{u} \times \textbf{B}\right)\right]
\end{alignat}
%
%
In the above equations, $\mathbf{u}$ is fluid velocity, $p$ is pressure, $\rho$ is density, $\mu$ is dynamic viscosity, $\gamma$ is surface tension coefficient, $\kappa$ is interface curvature, $\mathbf{n}$ is interface normal, $\delta$ is the delta function, $\mathbf{x}$ is the spatial location where the equation is solved, $\mathbf{x}_f$ is the position of the interface, $\mathbf{g}$ is acceleration due to gravity,  $\textbf{J}$ is electrical current density, $\textbf{B}$ is external magnetic field, $\Phi$ is electric potential, and $\sigma$ is electrical conductivity.

$\textbf{F}_L$ is the Lorentz force and calculated as $\textbf{F}_L = \textbf{J} \times \textbf{B}$, the electrical current density $\textbf{J}$ is calculated from Ohm's law as given by eq. \eqref{eqn:emf_ohms} and the electric potential ($\Phi$) is calculated from the eq. \eqref{eqn:emf_phi}. We also solve an equation to enforce divergence free condition $(\nabla \cdot \textbf{J} = 0)$ for the current density and maintain the conservative properties of charge. 

The liquid and gas distribution is distinguished by liquid volume fraction $(\alpha)$. We solve an advection equation of $\alpha$ given by eq. \eqref{eqn:vof_eqn} to capture the movement of interface between two fluids.
\begin{alignat}{1}
	\label{eqn:vof_eqn}
	\frac{\partial \alpha}{\partial t} + \textbf{u} \cdot \nabla \alpha = 0 
\end{alignat}
We use the linear weighting of the liquid volume fraction to calculate the mixture density, viscosity and electrical conductivity as,
%
\begin{alignat}{2}
	\label{eqn:density_viscosity}
	\rho &= \alpha \rho_l+ (1 - \alpha)\rho_g \\
	\mu  &= \alpha \mu_l + (1 - \alpha)\mu_g \\
	\sigma &= \alpha \sigma_l+ (1 - \alpha)\sigma_g
\end{alignat}
The subscripts $l$ and $g$ denote gas and liquid phases respectively. 

\subsection*{Solution procedure}

The accuracy and speed of a multiphase algorithm is dependent on methods to track the interface, model surface tension force, reduce momentum imbalance across the interface, handle property discontinuity, etc. In our current numerical procedure, we have used the geometry construction method \cite{Rider1998} to represent the interface inside a cell and to calculate volume fraction fluxes at the cell faces. The interface normals are calculated from $\alpha$ using second-order central differencing scheme for derivatives. The second order operator split method of \cite{Noh1976,Li1995,Ashgriz1991} is used for solution of the VOF equation. Here, the advection along one direction (x, y or z) is done as a first step then the advection along remaining two directions are performed. We rotate the order of advection to reduce any directionally-biased errors introduced due to operator splitting approach. The advection term in the momentum equation (eq. \eqref{eqn:momentum_equation})  is also calculated geometrically is ensure consistency between volume fraction and momentum advections. \cite{Kumar2016thesis,Kumar2015numerical} have provided more details about challenges and mitigation associated with implementation geometry construction method in three-dimensions.

The $\gamma \kappa \mathbf{n} \delta \left( \mathbf{x} - \mathbf{x}_f \right)$ term in eq. \eqref{eqn:momentum_equation} represents the surface tension force at the interface with $\gamma \kappa$ as the Laplace pressure jump. In our current algorithm, we have used the Sharp Surface Force (SSF) \citep{Francois2006,Wang2008,Kumar2019AJKFluids} method to model the surface tension term in the Navier-Stokes equation. Here, the surface tension term is written as a pressure gradient term given by, 
\begin{alignat}{1}
	\label{eqn:jump_condition}
	\gamma \kappa \mathbf{n} \delta \left( \mathbf{x} - \mathbf{x}_f \right)  = - \nabla \tilde{p} 
\end{alignat}
where $\tilde{p}$ is the pressure solely due to the surface tension force at the interface. We obtain a Poisson equation for $\tilde{p}$ from continuity and momentum equations,
\begin{align}
	\label{eqn:ppe_surface_tension}
	\nabla \cdot \left( \frac{\nabla \tilde{p}}{\rho} \right) = 0
\end{align}
The jump condition at the interface given by eq.\eqref{eqn:jump_condition} is used in the solution of eq.\eqref{eqn:ppe_surface_tension}. This ensures the exact difference in pressure at the interface due to the surface tension. Using this method the spurious velocities are seen to reduce to machine zero for a static bubble with exact analytical curvature and to very small values when the curvature is numerically computed. We have used the height function method of \cite{Rudman1998,Cummins2005} for curvature estimation. Admittedly, there are several methods to include the surface tension force in the Navier-stokes (Renardy and Renardy \cite{Renardy2002}, Sussman et al. \cite{Sussman2003,Sussman2007}, Gueyffier et al. \cite{Gueyffier1999}), however, in our experience the current method is more accurate for reduction of spurious velocities and relatively easier to implement on GPUs. \par

The momentum and continuity equations are discretized on a collocated Cartesian grid using a finite volume method. The terms in the  momentum equation  are integrated with a second-order accurate in time and space using Adams–Bashforth time advancement. The convection term($\nabla \cdot \rho u u$) is computed geometrically with linear interpolation for the face velocity. The pressure gradient term is written at cell faces and in the form of $\frac{\nabla p}{\rho}$. Coupling the pressure gradient and density together eliminates any ambiguity in density interpolation and leads to a robust method. The viscous term is computed by second order central differencing scheme. Linear interpolation is used for face densities and viscosities.

The electromagnetic equations given by eq. \eqref{eqn:emf_ohms} and \eqref{eqn:emf_phi} are solved on the same grid as used for continuity and momentum equations. The magnetic field strength $(\textbf{B})$, electric potential $(\Phi)$, current density $(\textbf{J})$ and thermal conductivity $(\sigma)$ are stored at the cell centers and calculated at cell faces using linear averaging scheme. 

We solve three Poisson equations: pressure $(p)$, surface tension pressure $(\tilde{p})$ and electric potential $(\Phi)$ in this numerical approach. These equations are efficiently solved using a geometric multigrid accelerated red black SOR relaxation scheme. 

All boundaries of the domain are considered wall, therefore no-slip, no-penetration and non-conducting boundary conditions are applied at them.

\begin{alignat}{1}
	u_{wall} = v_{wall} = w_{wall} = 0 \\
	\left(\frac{\partial p}{\partial \hat{n}}\right)_{wall} = 	\left(\frac{\partial \tilde{p}}{\partial \hat{n}}\right)_{wall} = 0 \\
	J_{wall} = J_{wall} = J_{wall} = 0 \\ 
	 \left(\frac{\partial \Phi}{\partial \hat{n}}\right)_{wall} = 0
\end{alignat} 
We have implemented this algorithm to run on multiple graphics processing units (GPU). We have used the CUDA-Fortran platform supported by the Portland group (PGI) Fortran compilers. Previously, we have validated the multiple GPU implementation \citep{Kumar2015numerical,Vanka2016Single,Kumar2016thesis} to study confinement effects on bubble dynamics \citep{Kumar2015confinement}, two-phase flows at T-juctions \citep{horwitz2012simulations,horwitz2013simulations,kumar2013three}, bubble dynamics in non-Newtonain fluids \citep{ Kumar2015non_newtonian,Kumar2019AJKFluids}, droplet dynamics in square duct \citep{Horwitz2014_lbm,Horwitz2019AJKFluids}, Argon bubble rising in liquid steel \citep{Vanka2015APS_DFD,Jin2016mhd_bubble,Kumar2022APS_DFD} and turbulent bubbly flow \citep{Vanka2016APS_DFD,kumar2021_bubbly_flow}. \par
From analysis of the governing equations, we have identified that the dynamics of bubbles rising in a confined conducting liquid medium depends on these non-dimensional parameters. 
\begin{alignat}{1}
	\text{Bond number} &= Bo = \rho_l gd^2/\gamma \\
	\text{Morton number} &= Mo = g \mu_{l}^4/\rho_l \gamma^3 \\
	\text{Hartmann number} &= Ha = B d \sqrt{\sigma/\mu_l} \\
	\text{Confinement ratio} &= C_r = W/d \\
	\text{center to center distance} &= h_d = h/d
\end{alignat}

\subsection*{Grid refinement study}
\label{subsec:grid_refinement}

We have conducted a systematic grid refinement study to understand the impact of grid size on the bubbles' dynamics. For this grid refinement study, we have considered three grids namely: coarse ($\Delta x = d/16: 64 \times 64 \times 384$), fine ($\Delta x = d/32: 128 \times 128 \times 768$), and finer ($\Delta x = d/64: 256 \times 256 \times 1536$), and have analyzed the bubbles' rise velocity and shapes for both leading and trailing bubbles. 
\begin{figure}[h]
	\begin{center}
		\begin{subfigure}[b]{0.48\textwidth}
			\includegraphics[width=1\textwidth,trim=4 4 4 4,clip]{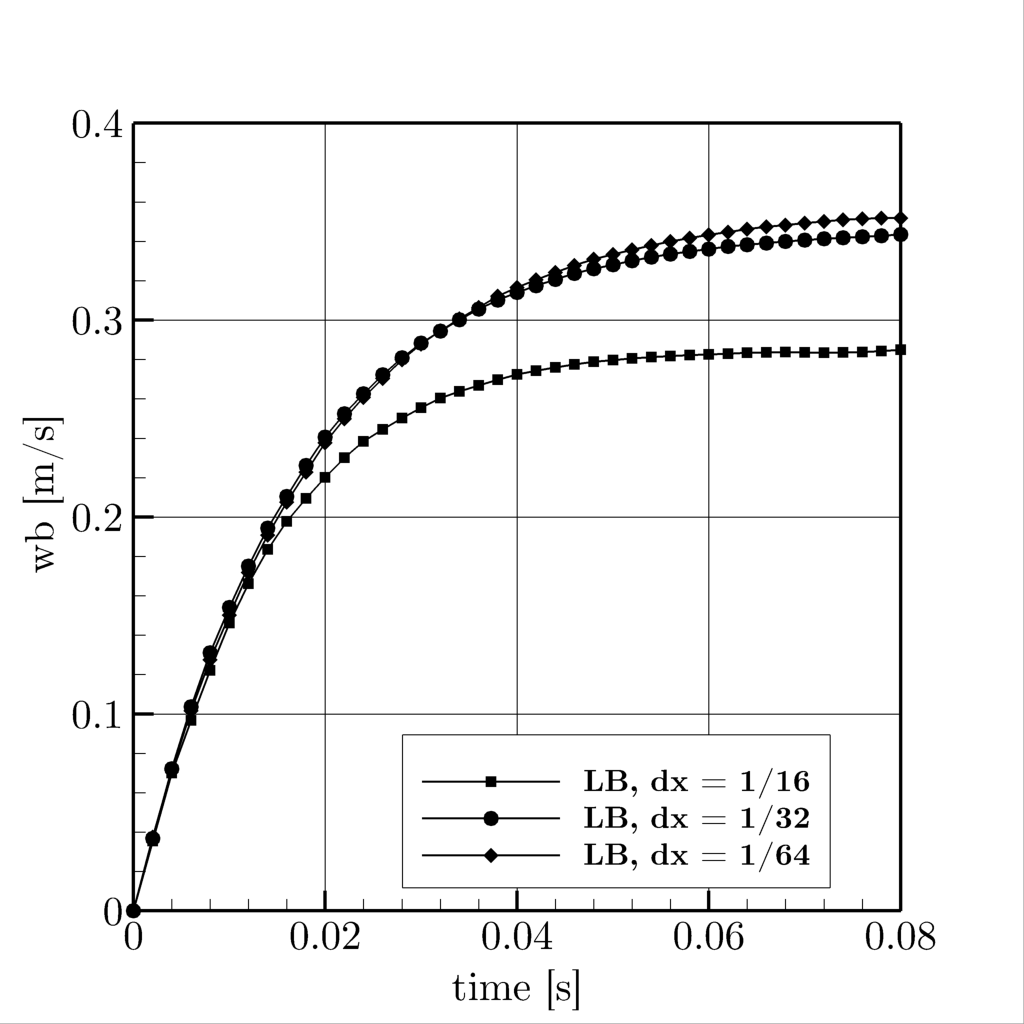}
			\caption{Rise velocity of leading bubble}
			\label{fig:gr_rv_LB}
		\end{subfigure} %
		\begin{subfigure}[b]{0.48\textwidth}
			\includegraphics[width=1\textwidth,trim=4 4 4 4,clip]{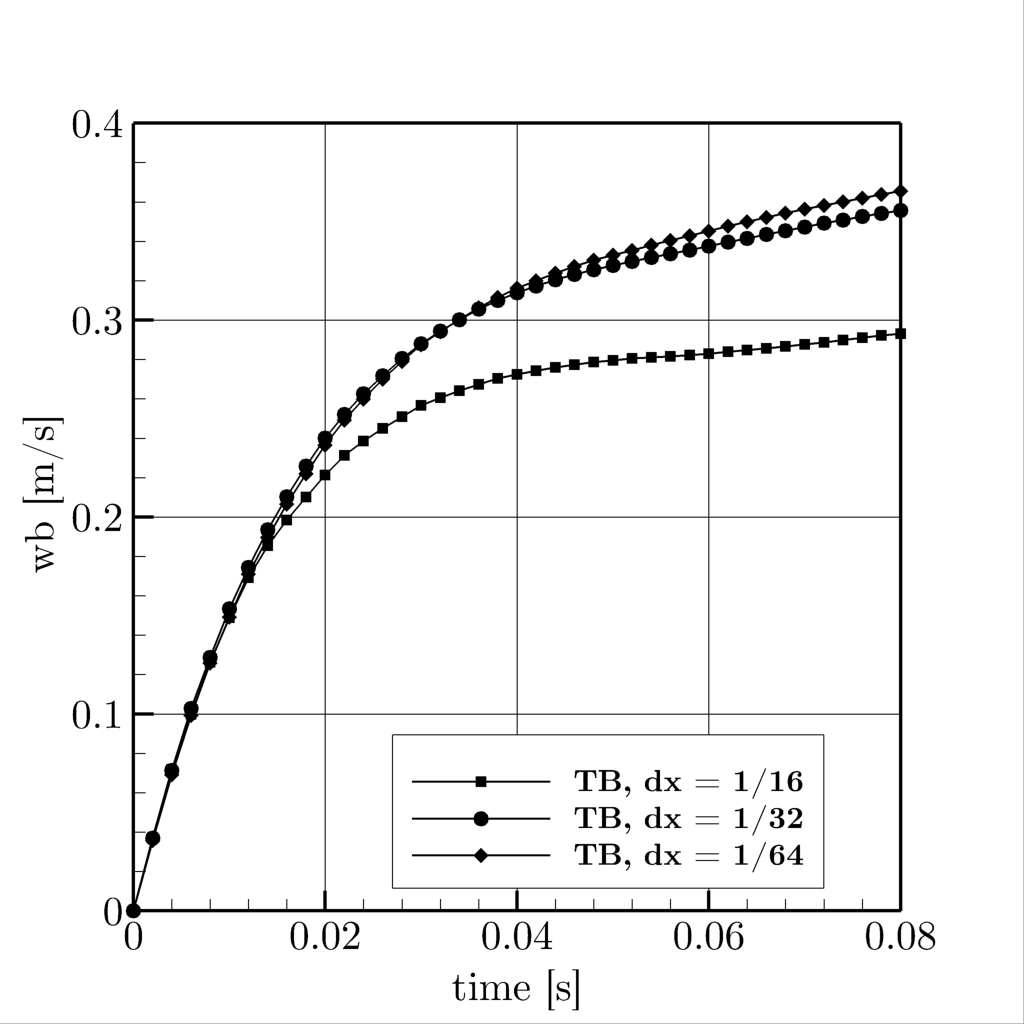}
			\caption{Rise velocity of trailing bubble}
			\label{fig:gr_rv_TB}
		\end{subfigure} %
		\caption{Rise velocity of leading and trailing bubbles for three grid sizes}
		\label{fig:gr_rv}
	\end{center}
\end{figure}

Figure \ref{fig:gr_rv} shows the rise velocities of the leading and trailing bubbles for three different grids. It can be noticed from the figure that both bubbles initially accelerate with faster rate then slows down to reach a steady state. For the total simulation time considered in this study, we notice that the leading bubble reach a steady state, however trailing bubble continues to accelerate. Further, we can observe that rise velocities of both leading and trailing bubbles for three grids are almost identical up to 15 ms and diverges afterwards. The velocities predicted on coarse grid is the lowest and increases with grid refinement. For leading bubble, rise velocities are 0.2848, 0.3434, and 0.3581 m/s at $t=0.08$ ms, for $\delta x = 1/16$, 1/32, and 1/64 grids respectively. The difference in rise velocities between finer and fine meshes is 4.1\% and between finer and coarse meshes is 20.5\%. Similarly, for trailing bubble, rise velocities are 0.2931, 0.3557, and 0.3654 m/s at $t=0.08$ ms, for $\delta x = 1/16$, 1/32, and 1/64 grids respectively. The difference in rise velocities between finer and fine meshes is 2.6\% and between finer and coarse meshes is 19.8\%. 

%
%

%
\begin{figure}[h]
	\begin{center}
		\begin{subfigure}[b]{0.48\textwidth}
			\includegraphics[width=1\textwidth,trim=4 4 4 4,clip]{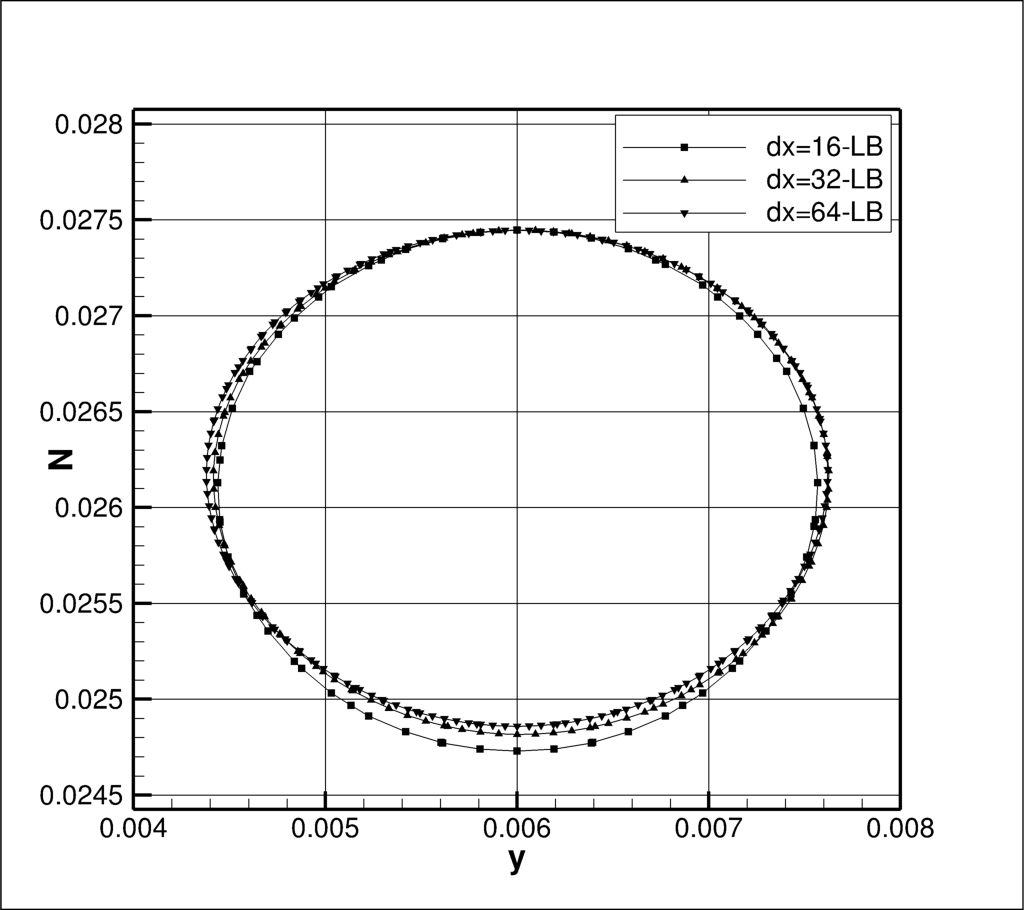}
			\caption{Shape of leading bubble}
			\label{fig:gr_shape_LB}
		\end{subfigure} %
		\begin{subfigure}[b]{0.48\textwidth}
			\includegraphics[width=1\textwidth,trim=4 4 4 4,clip]{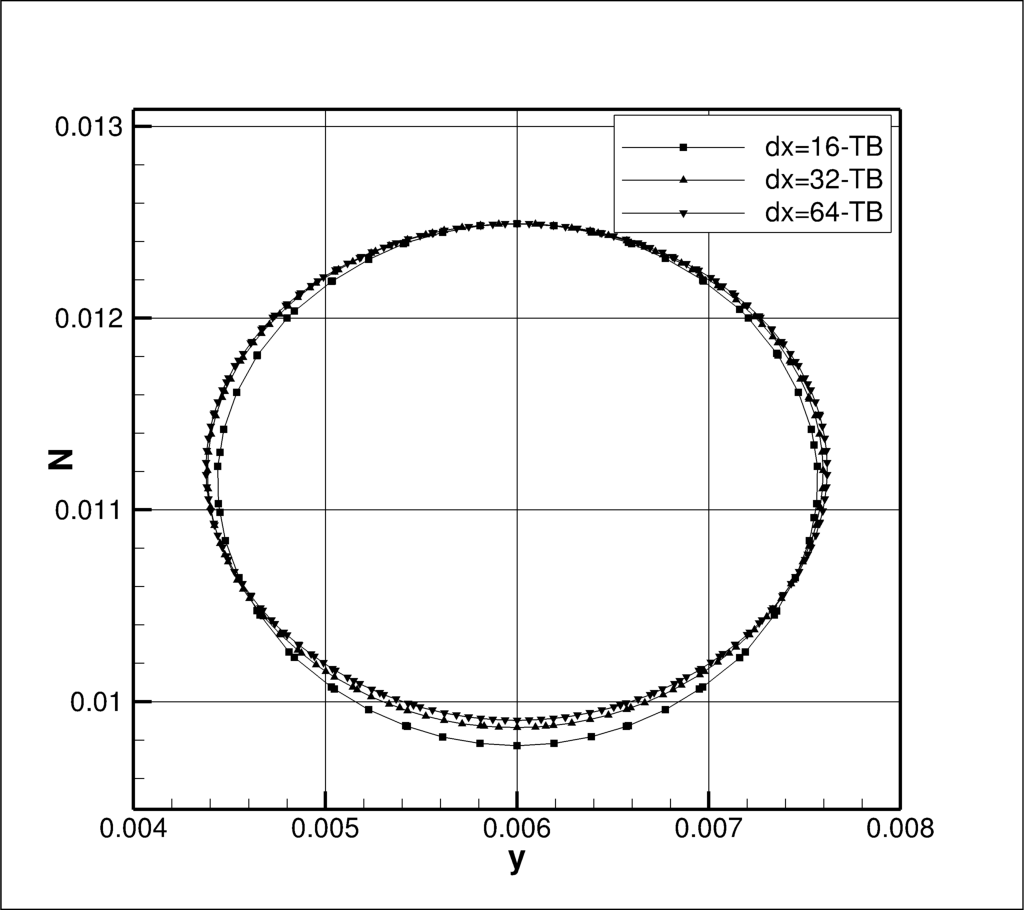}
			\caption{Shape of trailing bubble}
			\label{fig:gr_shape_TB}
		\end{subfigure} %
		\caption{Shape of leading and trailing bubbles for three grid sizes}
		\label{fig:gr_shape}
	\end{center}
\end{figure}

Figure \ref{fig:gr_shape} shows the bubble shape on the central plane of the domain at a time when they leading bubbles have reached steady state. For better comparison of the results, we have aligned bubble tops for all three grids. It can be noticed that bubbles become flatter as we refined the mesh, and $\Delta x = 1/16$ grid predicts largest height and smallest width for both leading and trailing bubbles. Both height and width converged as we refined the grid to $\Delta = 1/32$ and $1/64$ with very small difference between fine and finer grids.

From rise velocity and shape comparisons for three grids, we decided to use $\Delta x = 1/32$ as a suitable grid that can give accurate result with moderate computational cost. The results discussed from here onward will be on fine ($\Delta x = d/32: 128 \times 128 \times 768$) grid.

\section{RESULTS AND DISCUSSION}
\label{sec:results}
We now present the results of a study to analyze the effects of transverse magnetic field on the bubble deformation, rise velocity, rise path, and vortex shedding. We have considered bubbles of diameter $7~mm$ with the leading bubble placed 3 diameters above the trailing one, and have used two magnetic fields (0 and 0.2 T) to understand their effects. For these set of conditions, the key non-dimensional parameters are $Bo = 2.8$, $Mo = 1.2 \times 10^{-12}$, $C_r = 4$, $h_d = 3$ and $Ha = 14.9$. It should be noted that the bubbles are initially placed at the center of cross-section, but they may come closer to the wall during their ascend. Hence, instantaneous confinement ratio is likely to change. \par 

%
%
%
We first look at the transient shapes of the bubbles as they travel upward. According to the Grace diagram \cite{Grace1973}, an initially spherical bubble of $Bo = 2.8$ rising in a liquid medium of $Mo$ = $1.2 \times 10^{-12}$ deforms into a flatter ellipsoidal shape with minimum surface oscillations. In our current consideration, due to additional forces introduced by magnetic field, bubbles' dynamics (including shape and trajectory) are expected to be different. \par
\begin{figure}[H]
	\begin{center}  
		\begin{tabular}{| *1{>{\centering\arraybackslash}m{0.3in}} | *1{>{\centering\arraybackslash}m{0.3in}} | *5{>{\centering\arraybackslash}m{0.75in}|} @{}m{0pt}@{}}
			\cline{3-7}
			\multicolumn{1}{c}{} & \multicolumn{1}{c|}{} & \multicolumn{5}{c|}{time (ms)} & \\ [2ex]
			\hline
			$B$ &  & \textbf{0} & \textbf{80} & \textbf{160}  & \textbf{240}  & \textbf{320} & \\[2ex] 
			\hline
			\textbf{0} & LB & 
			\includegraphics[width=0.75in,trim=4 4 4 4,clip]{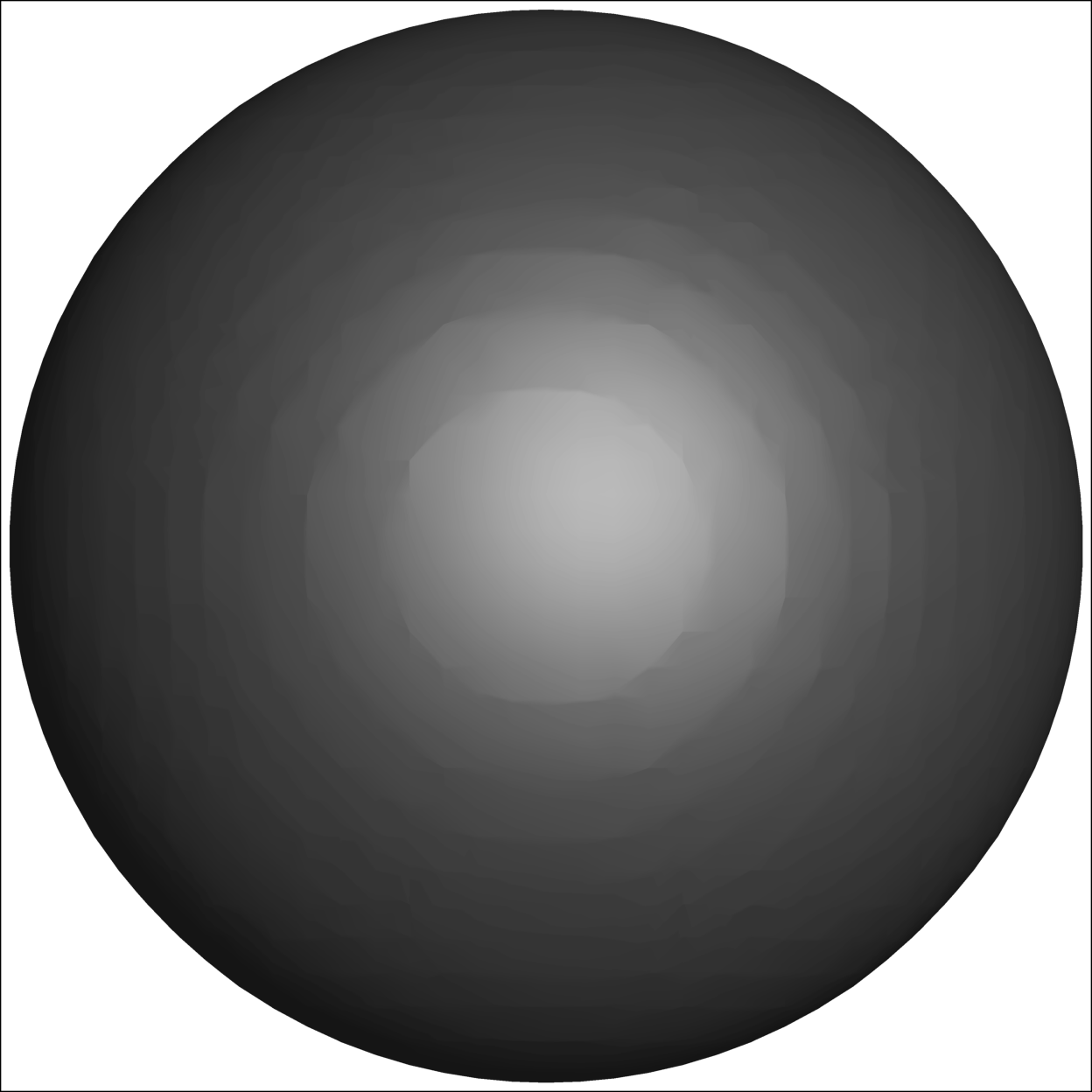} & \includegraphics[width=0.75in,trim=4 4 4 4,clip]{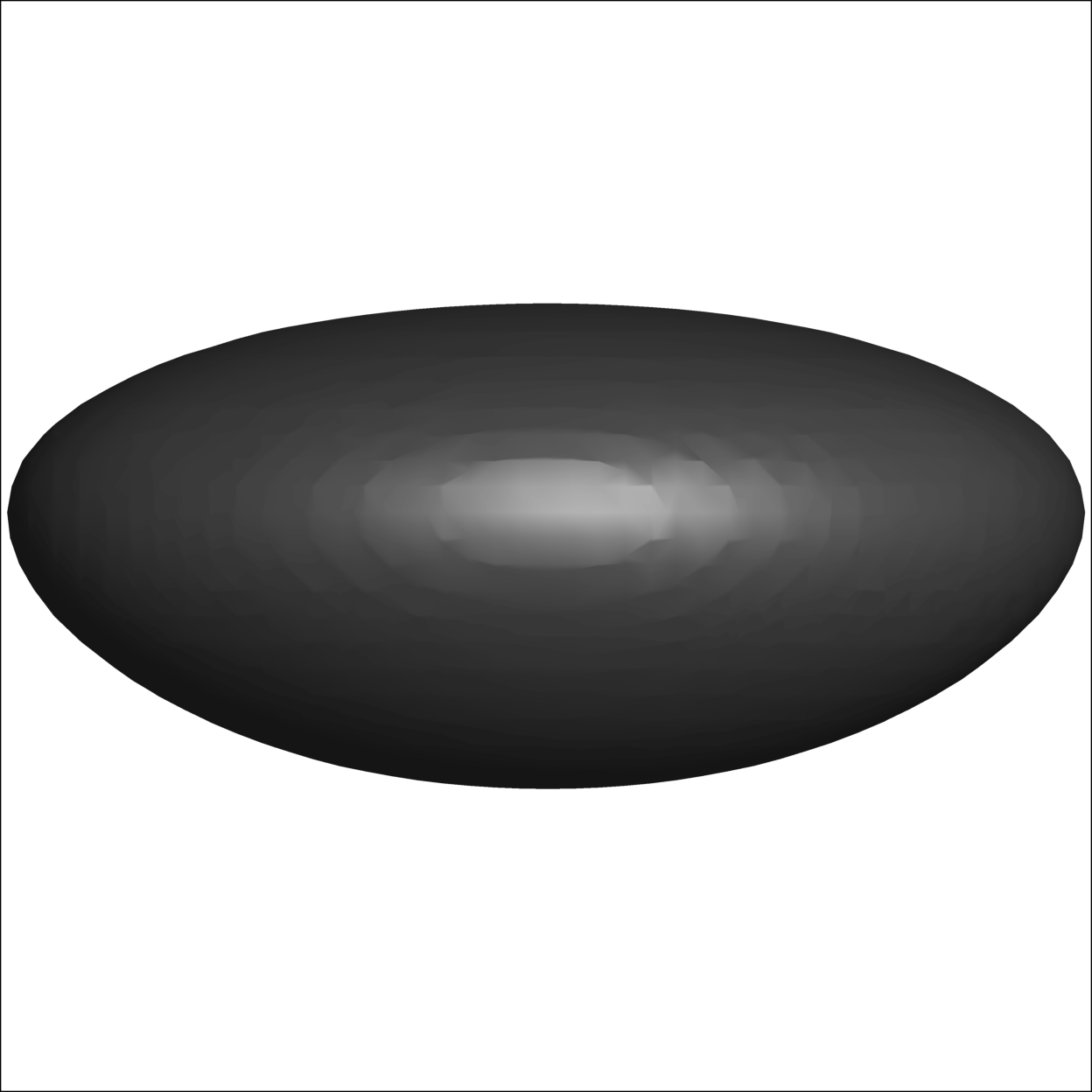} & \includegraphics[width=0.75in,trim=4 4 4 4,clip]{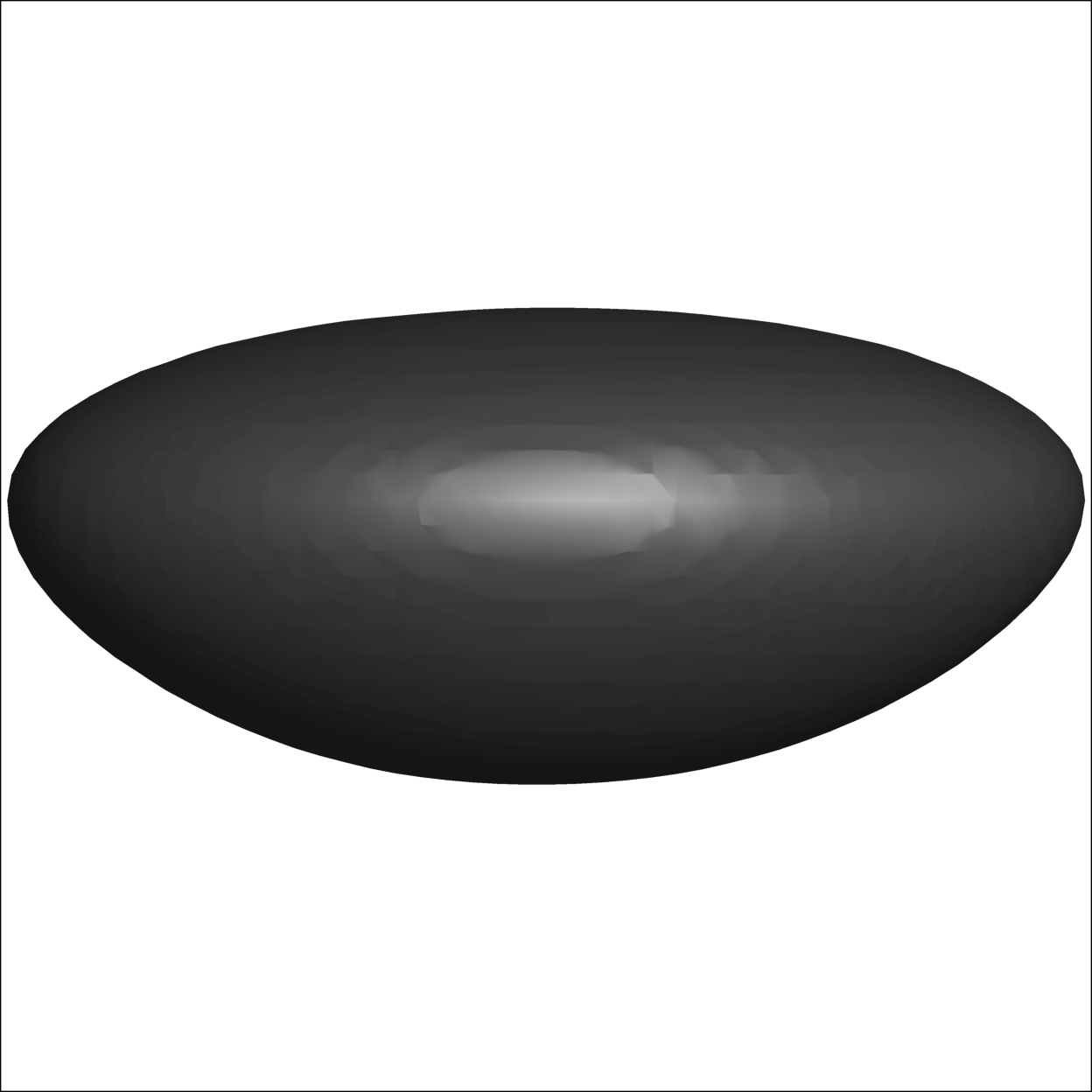} & \includegraphics[width=0.75in,trim=4 4 4 4,clip]{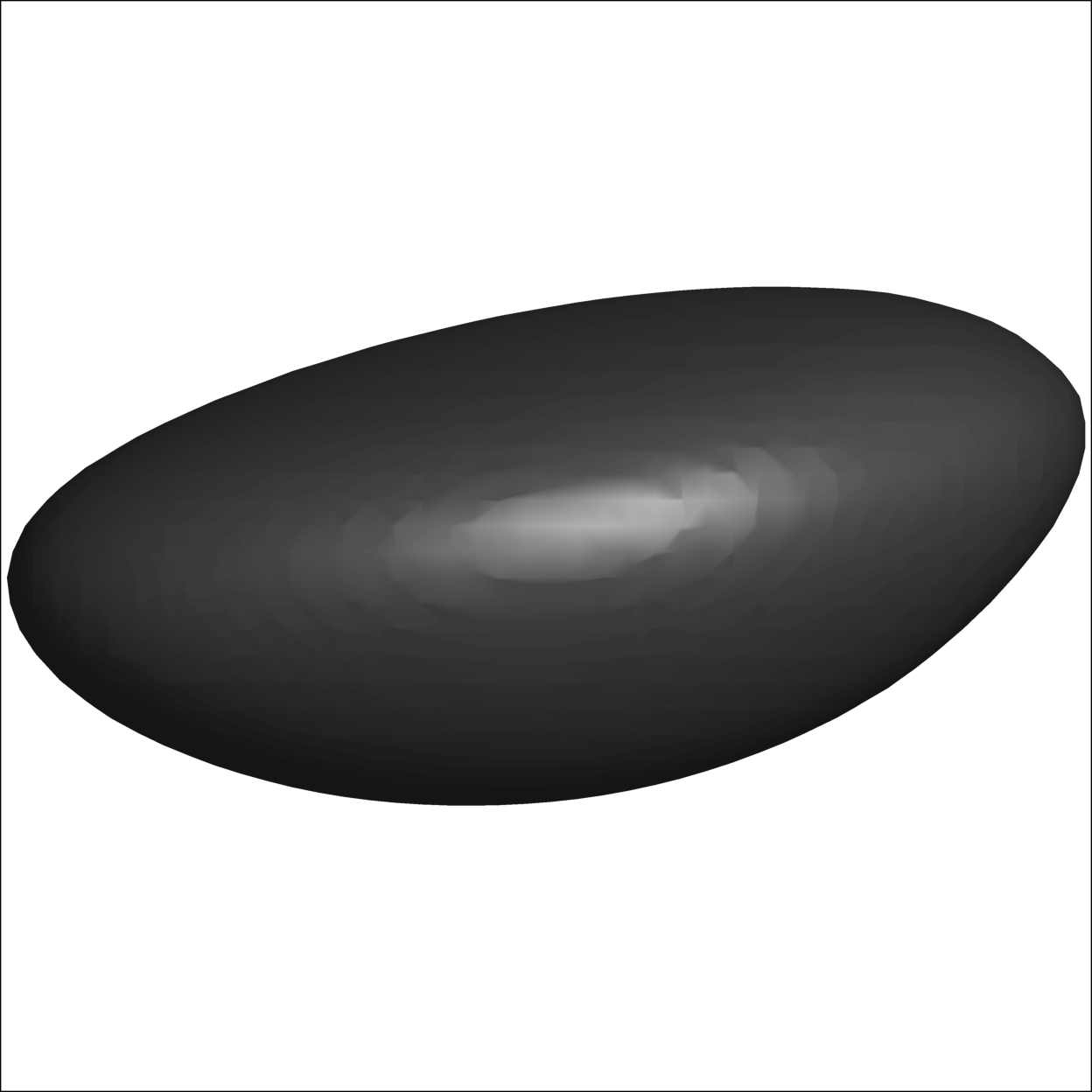} & \includegraphics[width=0.75in,trim=4 4 4 4,clip]{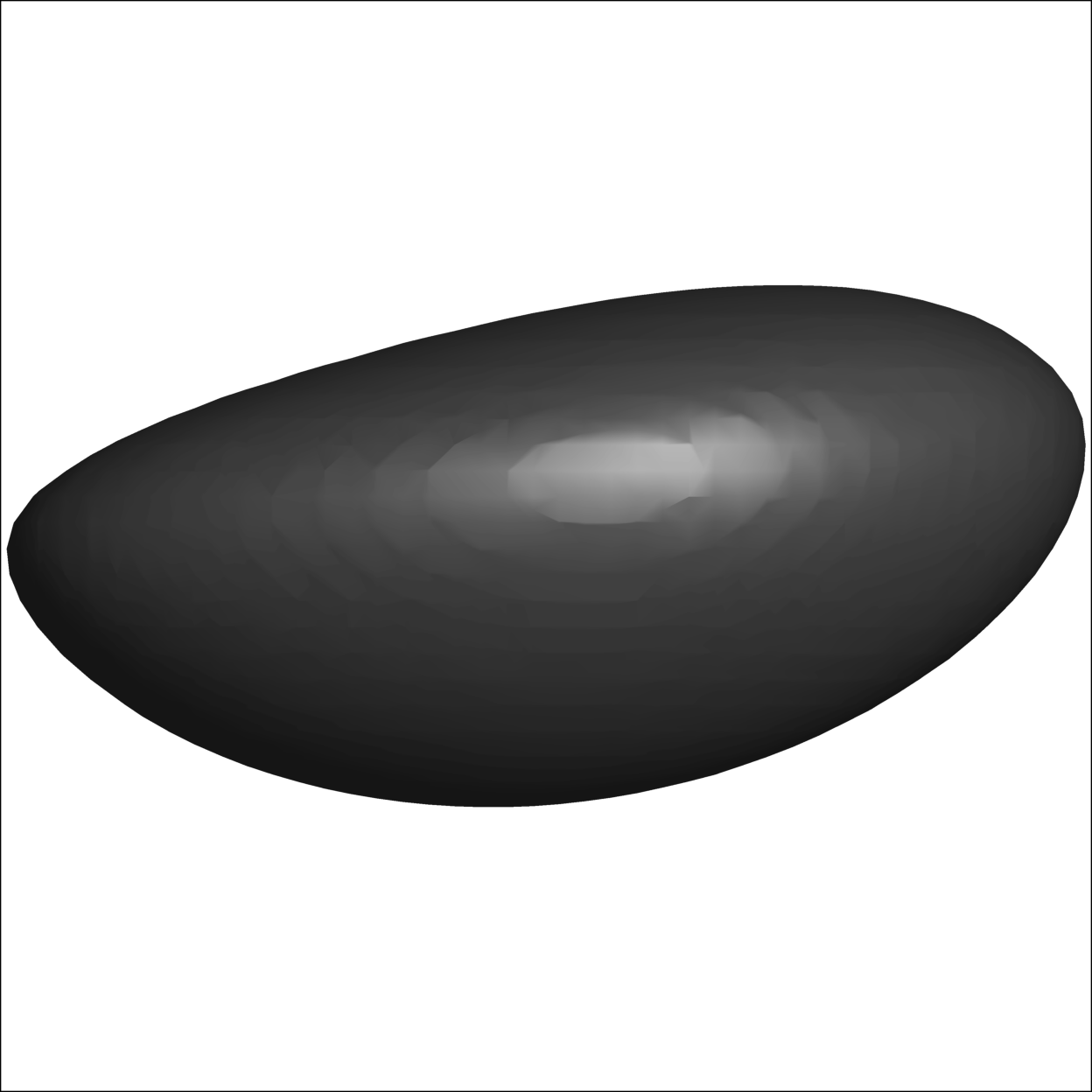} & \\[0ex]
			\hline
			\textbf{0} & TB &
			\includegraphics[width=0.75in,trim=4 4 4 4,clip]{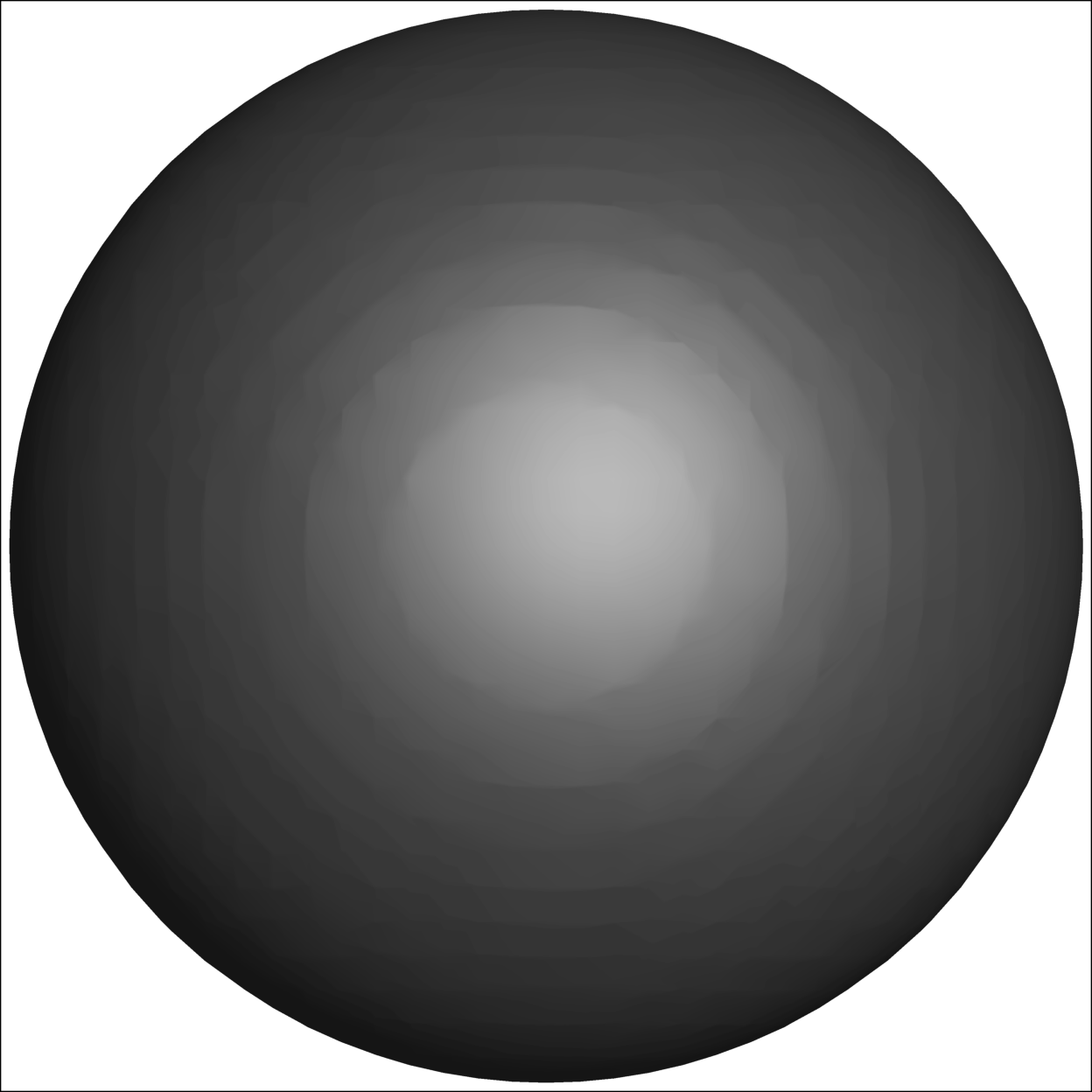} & \includegraphics[width=0.75in,trim=4 4 4 4,clip]{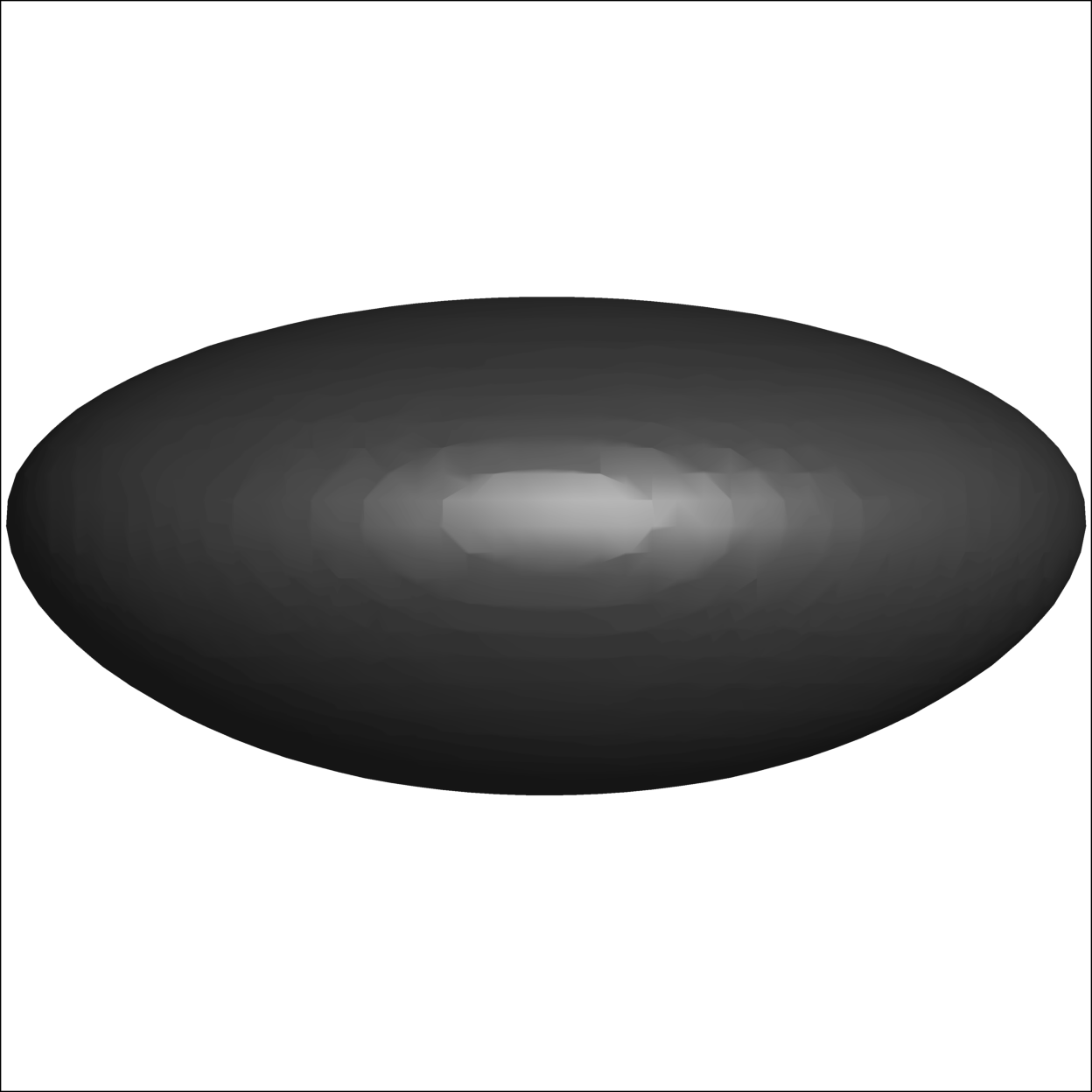} & \includegraphics[width=0.75in,trim=4 4 4 4,clip]{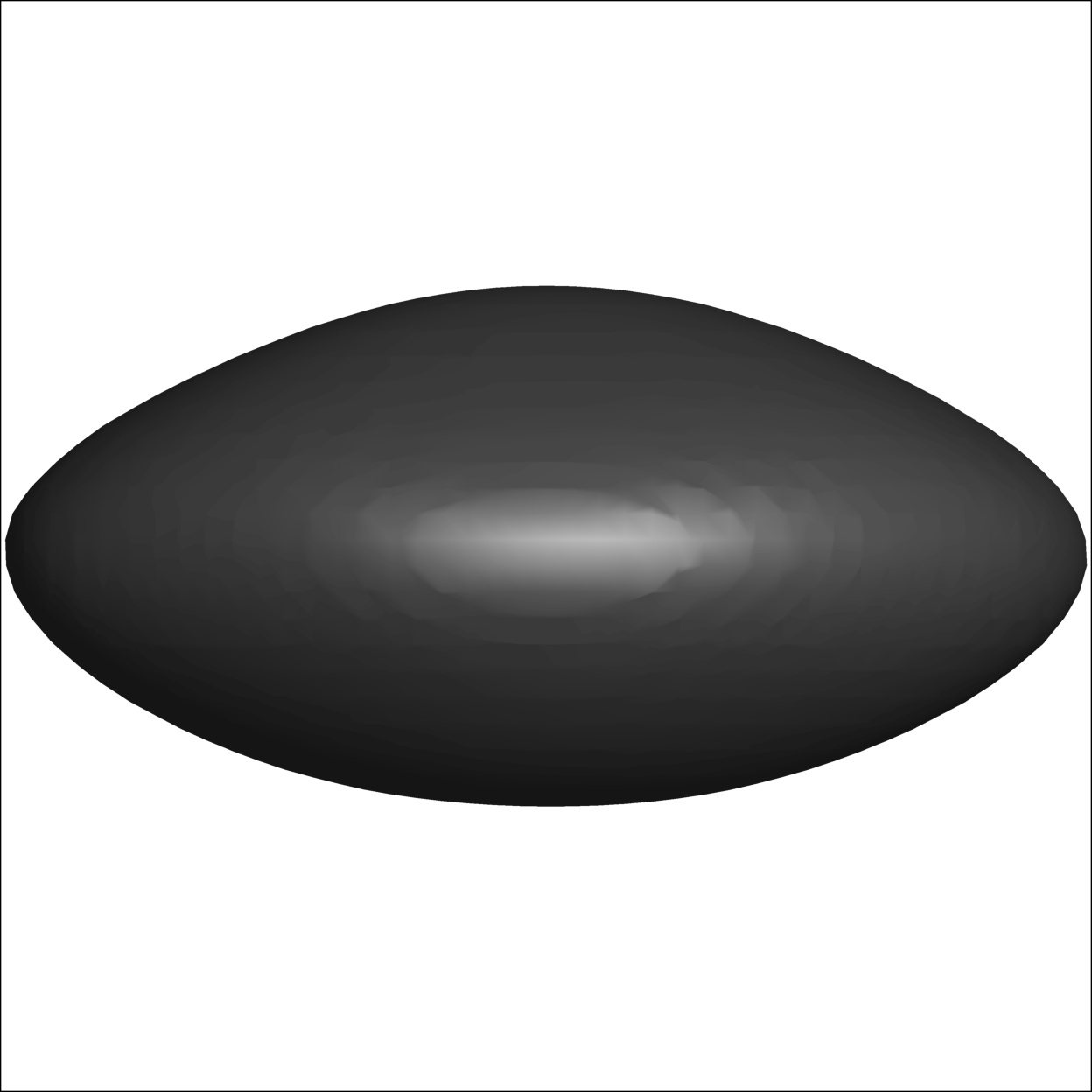} & \includegraphics[width=0.75in,trim=4 4 4 4,clip]{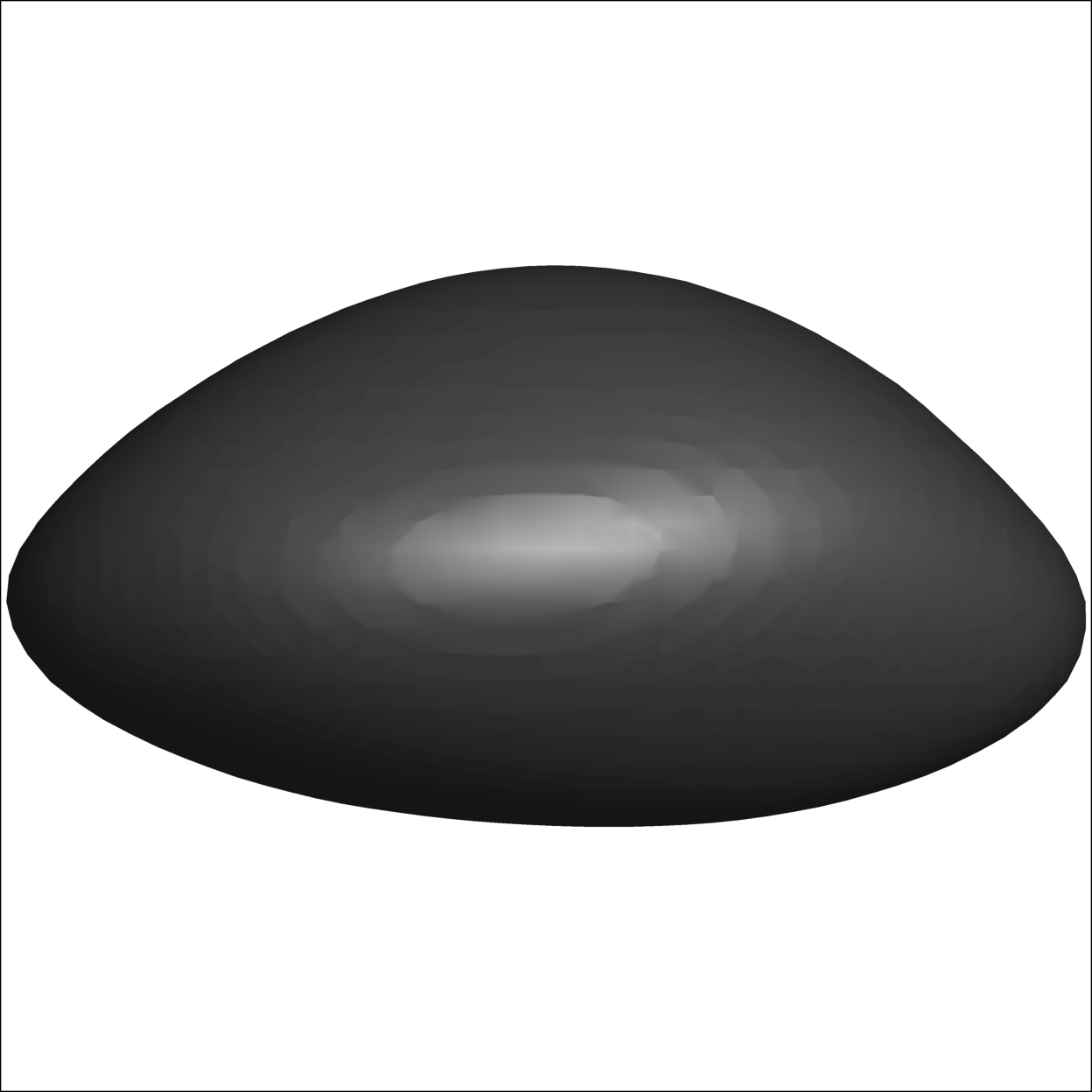} & \includegraphics[width=0.75in,trim=4 4 4 4,clip]{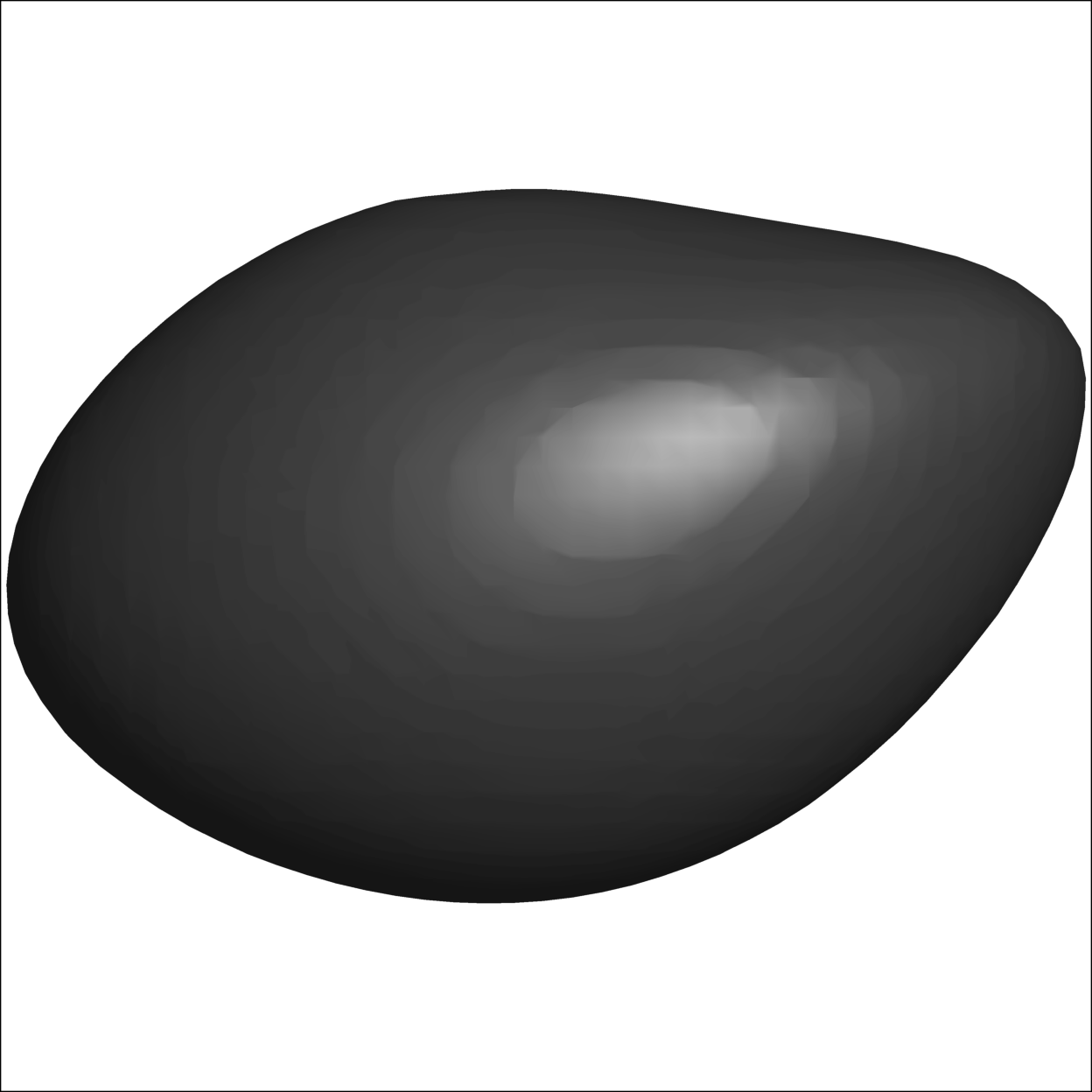} & \\[0ex]
			\hline
			\textbf{0.2} & LB &
			\includegraphics[width=0.75in,trim=4 4 4 4,clip]{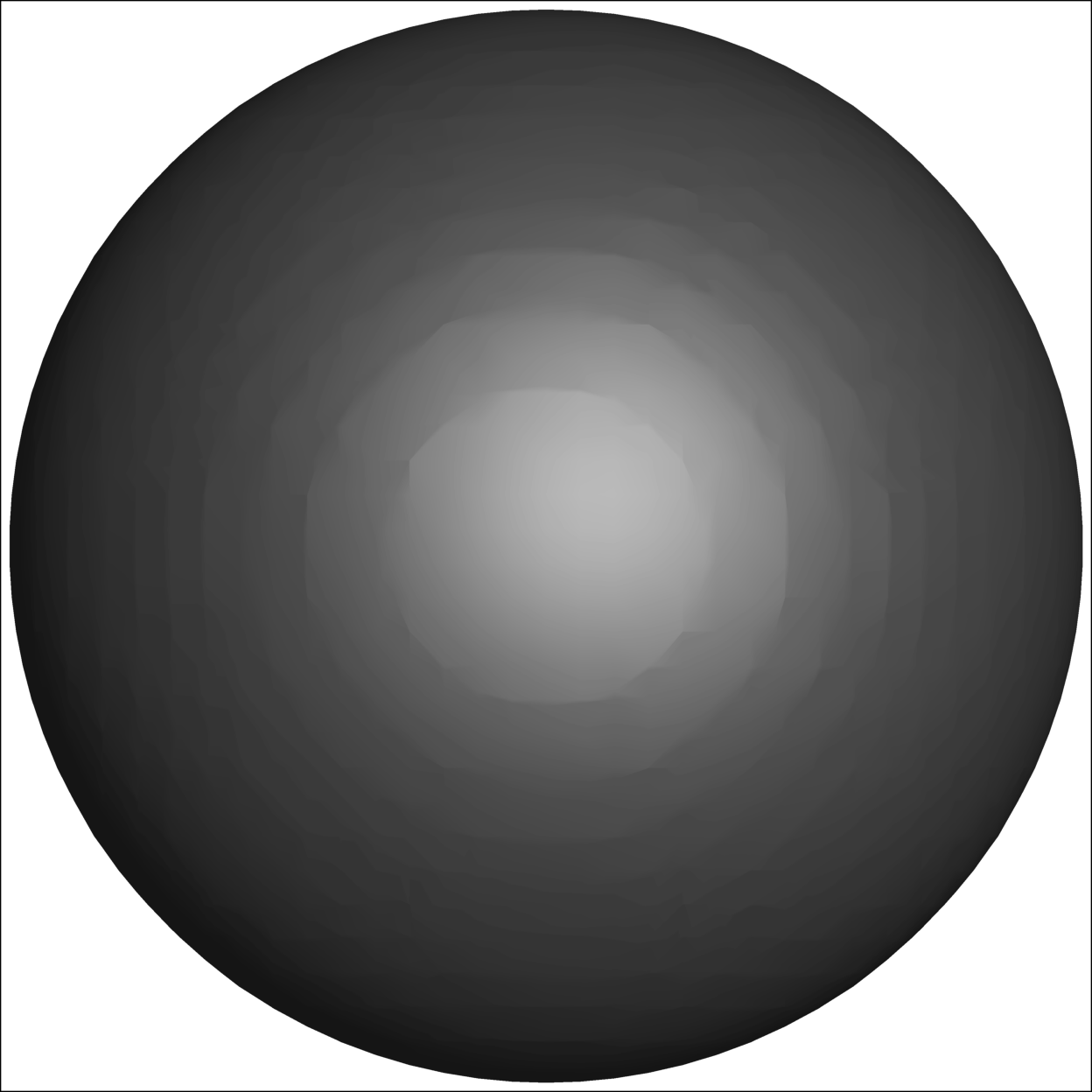} & \includegraphics[width=0.75in,trim=4 4 4 4,clip]{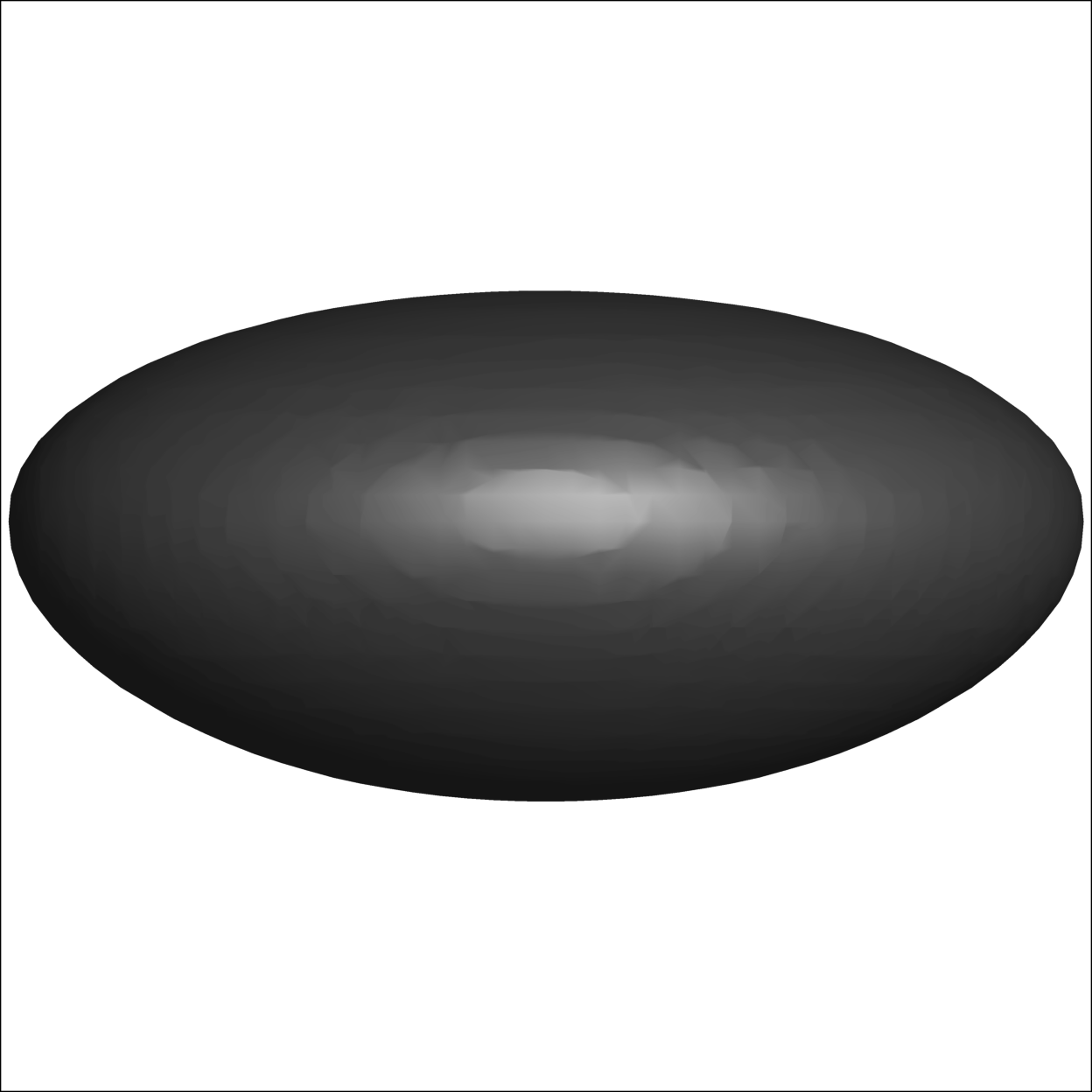} & \includegraphics[width=0.75in,trim=4 4 4 4,clip]{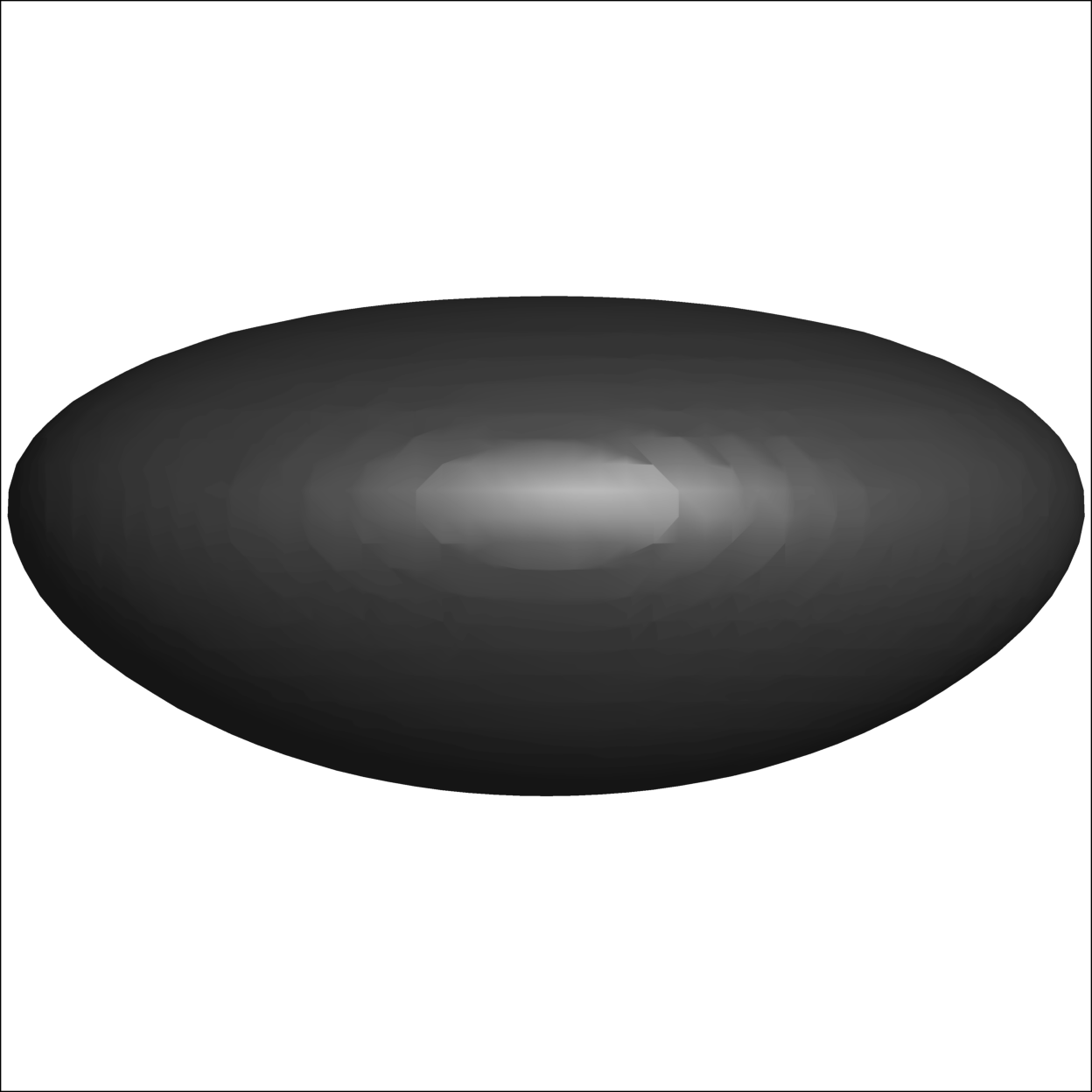} & \includegraphics[width=0.75in,trim=4 4 4 4,clip]{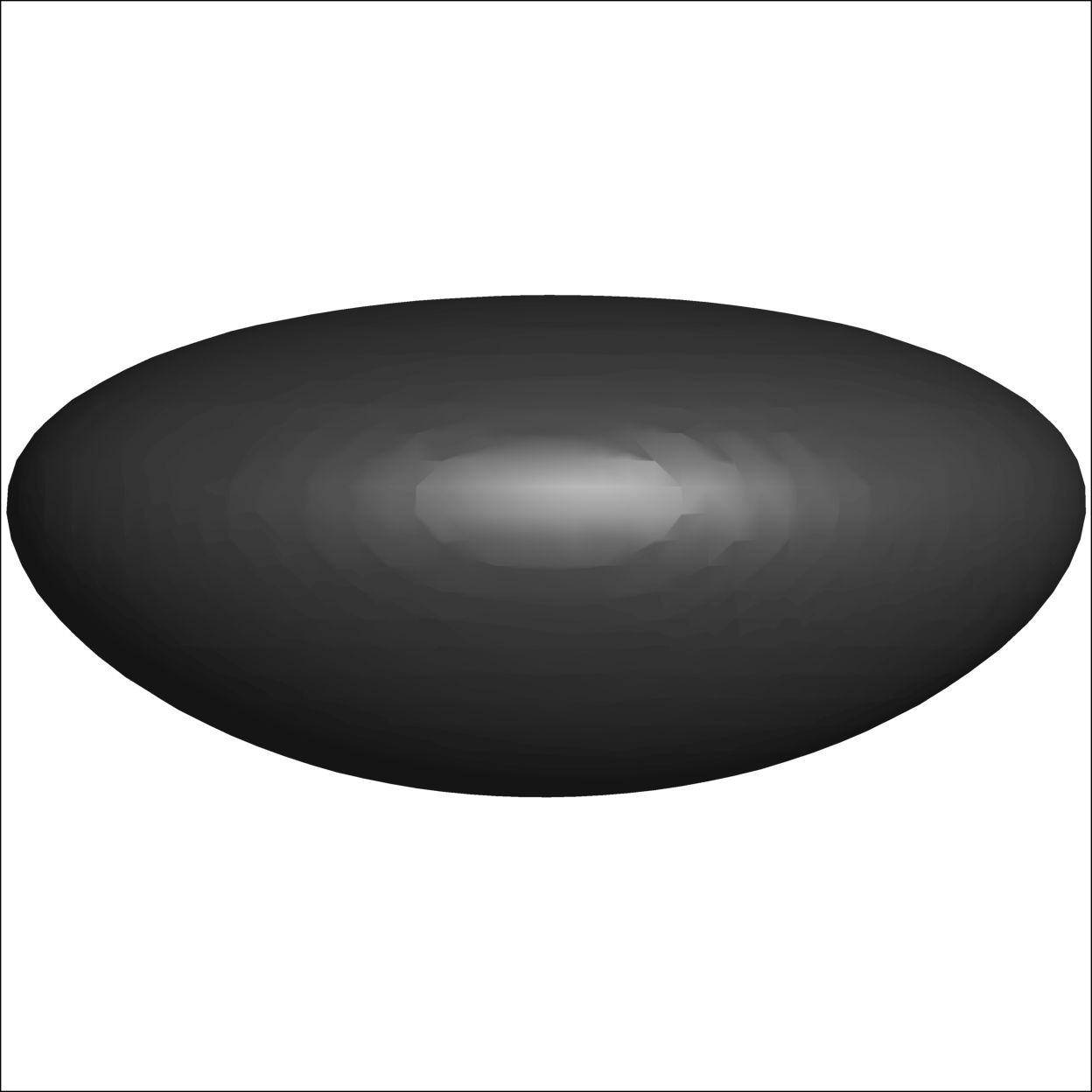} & \includegraphics[width=0.75in,trim=4 4 4 4,clip]{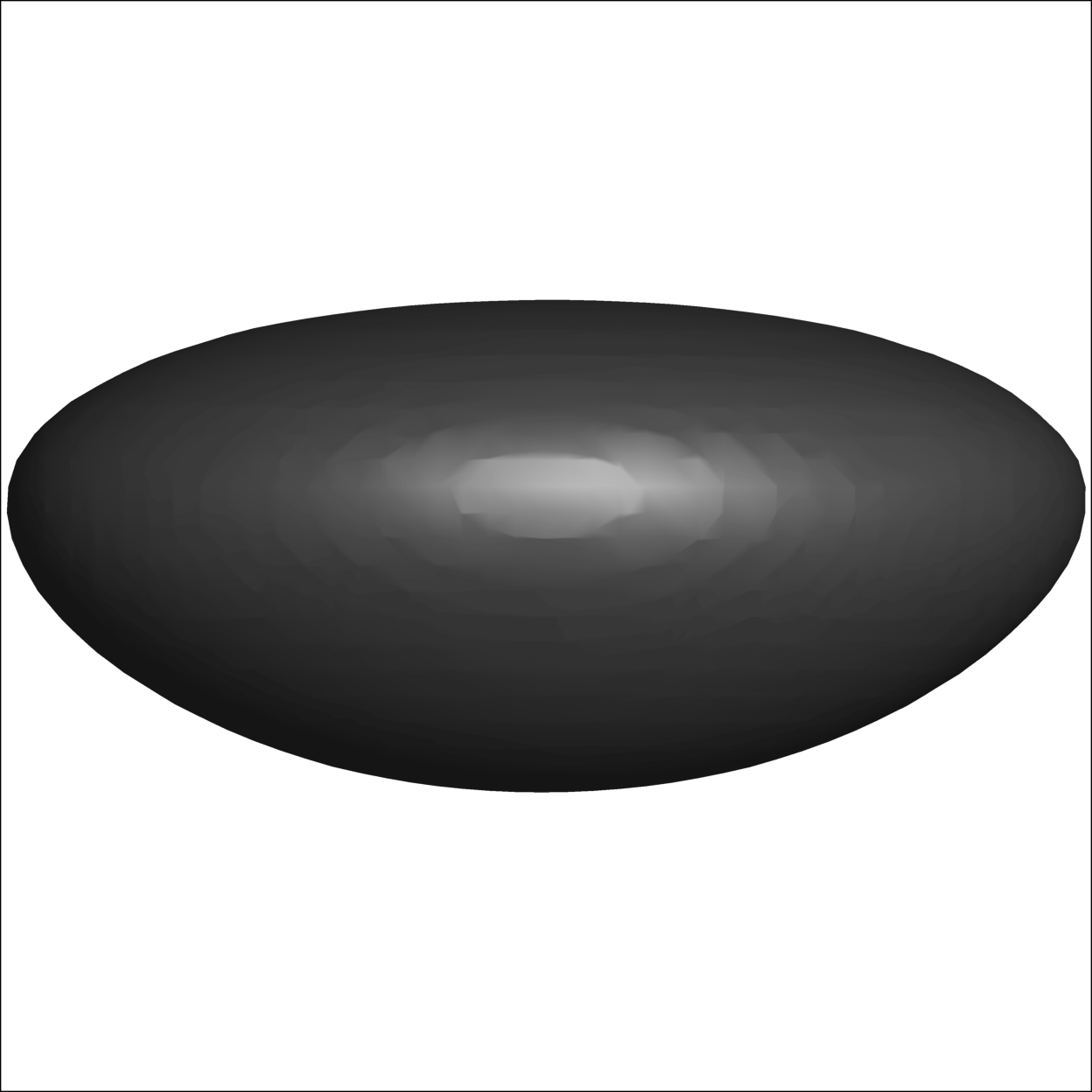} & \\[0ex]
			\hline
			\textbf{0.2} & TB &
			\includegraphics[width=0.75in,trim=4 4 4 4,clip]{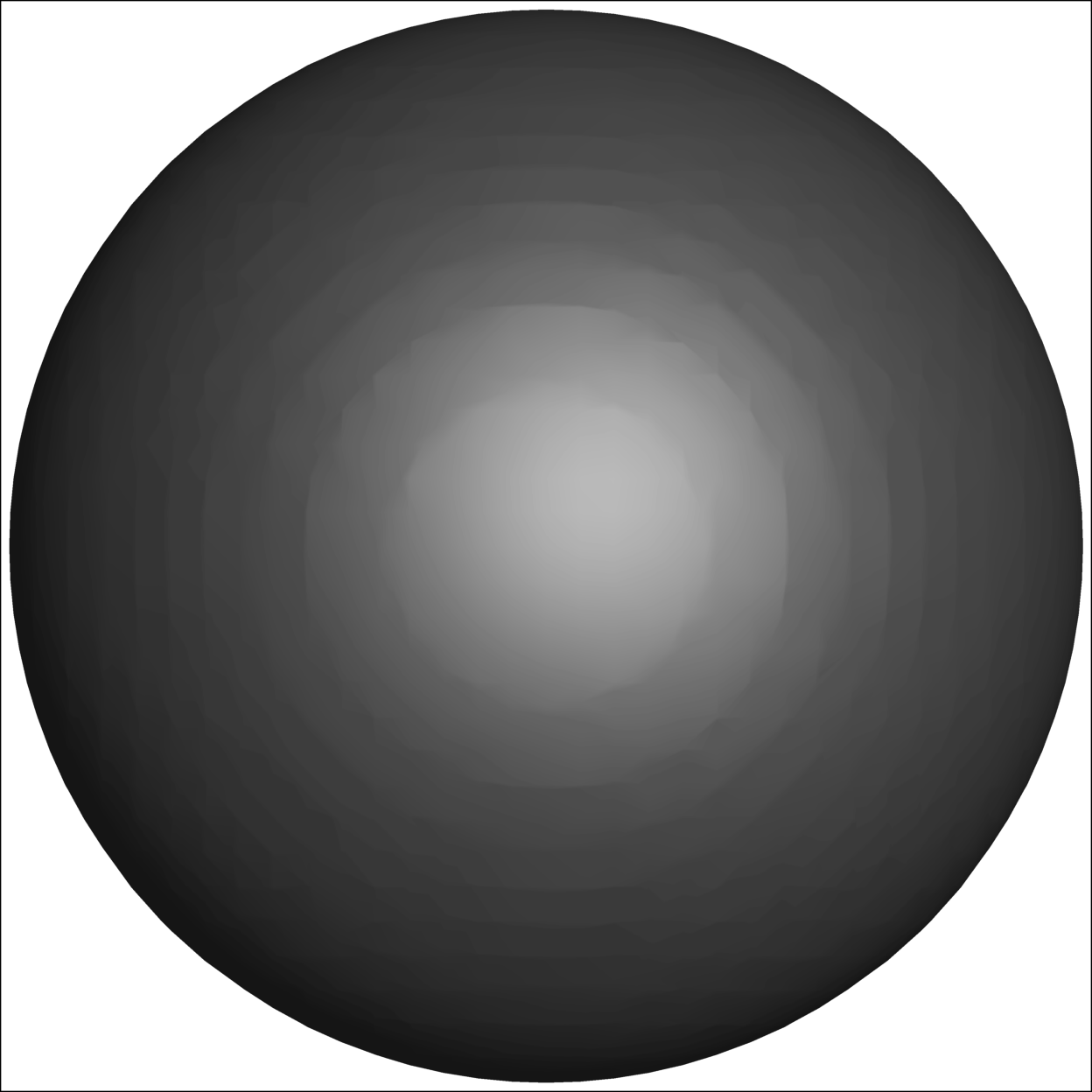} & \includegraphics[width=0.75in,trim=4 4 4 4,clip]{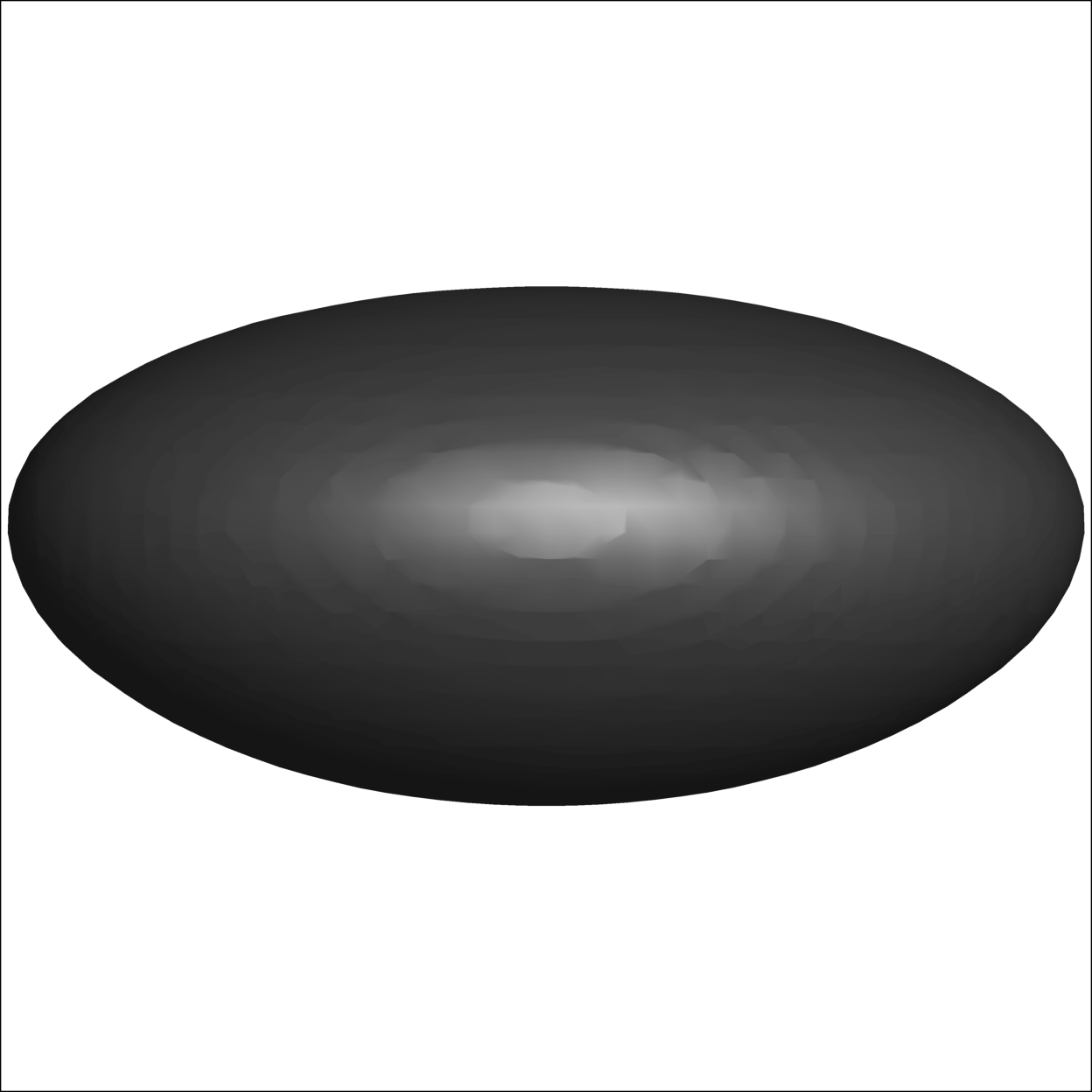} & \includegraphics[width=0.75in,trim=4 4 4 4,clip]{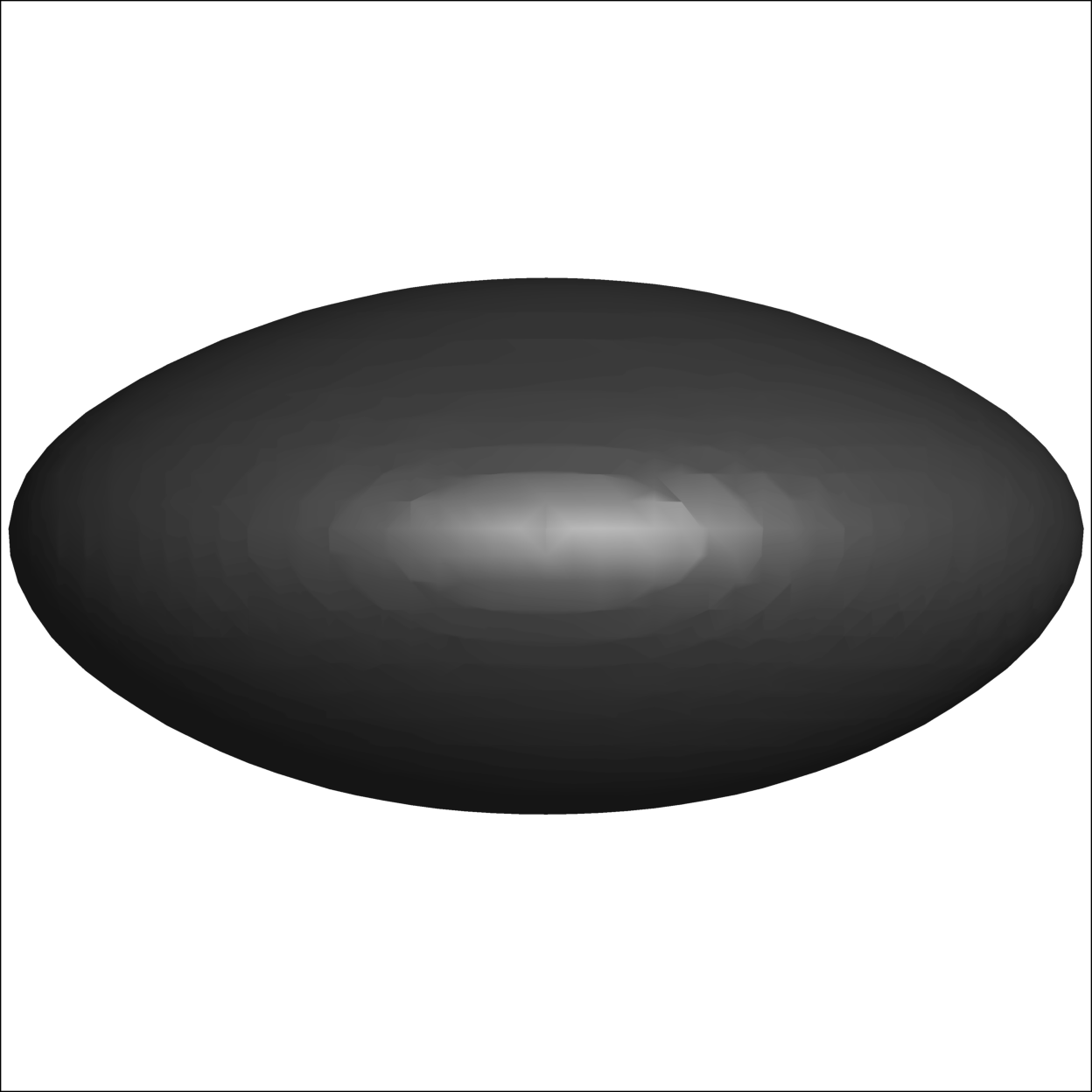} & \includegraphics[width=0.75in,trim=4 4 4 4,clip]{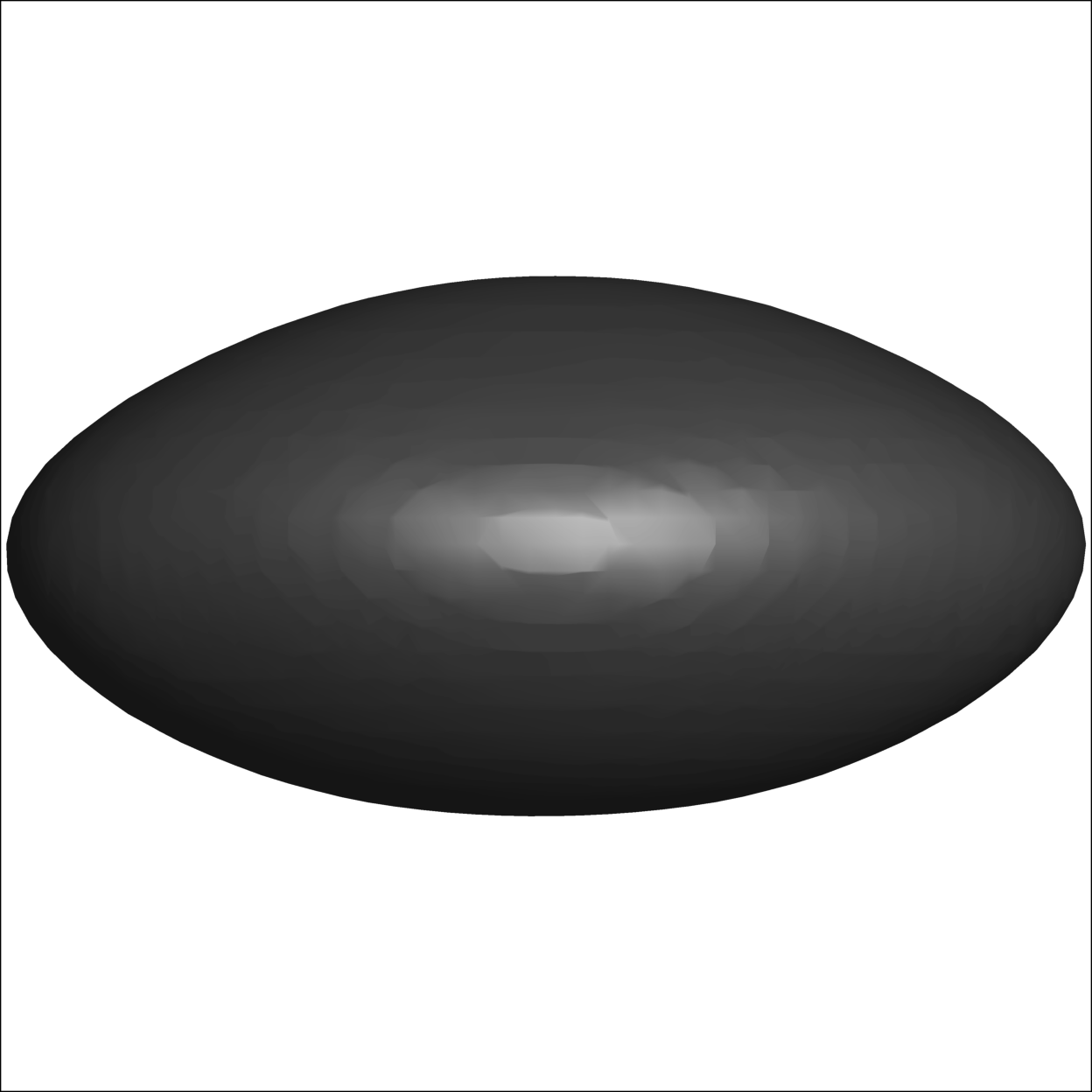} & \includegraphics[width=0.75in,trim=4 4 4 4,clip]{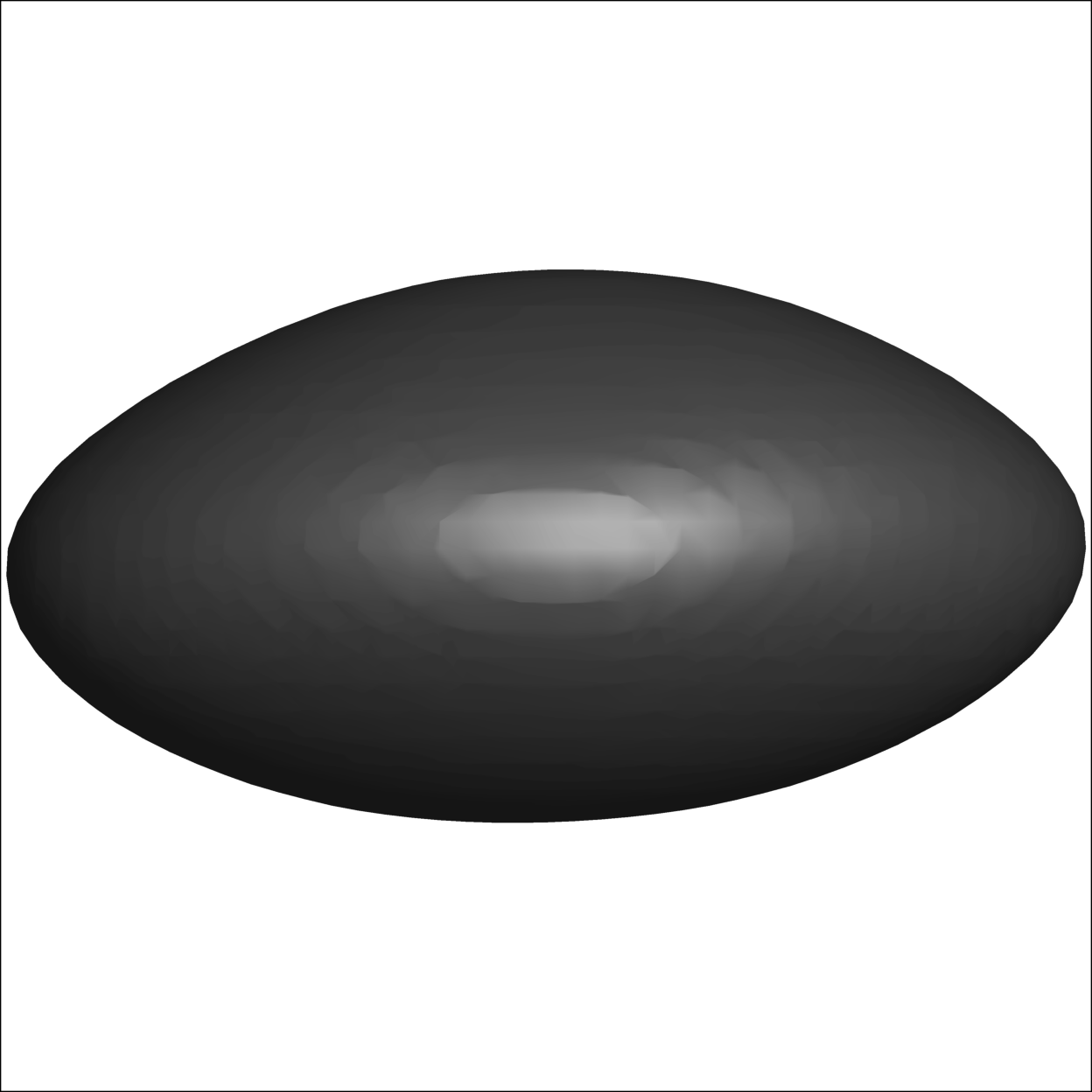} & \\[0ex]
			\hline
		\end{tabular}
		\caption{Transient three-dimensional front views of the bubbles for two magnetic fields}
		\label{fig:iso_surfaces_d_7mm_hd_3_front}
	\end{center}
\end{figure}
Figure \ref{fig:iso_surfaces_d_7mm_hd_3_front} shows the transient shapes of leading (LB) and trailing (TB) bubbles for $\textbf{B}$ = 0 and 0.2. In absence of magnetic field, the leading bubble deforms from a spherical shape to an ellipsoidal shape with asymmetrical top and bottom surfaces by 80 ms and remains ellipsoidal for remaining of the simulation time. We can notice that for up to 160 ms, bubble's major axes are horizontal and transitions to side-to-side oscillations (wobbling) between 160 - 240 ms. The trailing bubble also goes through similar transition from spherical to ellipsoidal by 80 ms, however it continues to deform afterwards. Shape oscillations of upper and lower surfaces: flatter-upper-surface (80 ms), rounder-upper-surface (160 ms), bulged-upper-surface (240 ms) to irregular upper surface afterwards.

In the presence of magnetic field, deformation of both leading and trailing bubbles are less pronounced than in the absence of magnetic field, however the trailing bubble still goes through more deformation than the leading one. We can notice that leading bubble reached a steady-state shape by 80 ms and did not go through shape oscillations or wobble motion. The trailing bubble, however, continued to deform after 80 ms and did not achieve a steady-state shape for the time considered here. 

%
%
\begin{figure}[H]
	\begin{center}
		\includegraphics[width=0.7\textwidth,trim=4 4 4 4,clip]{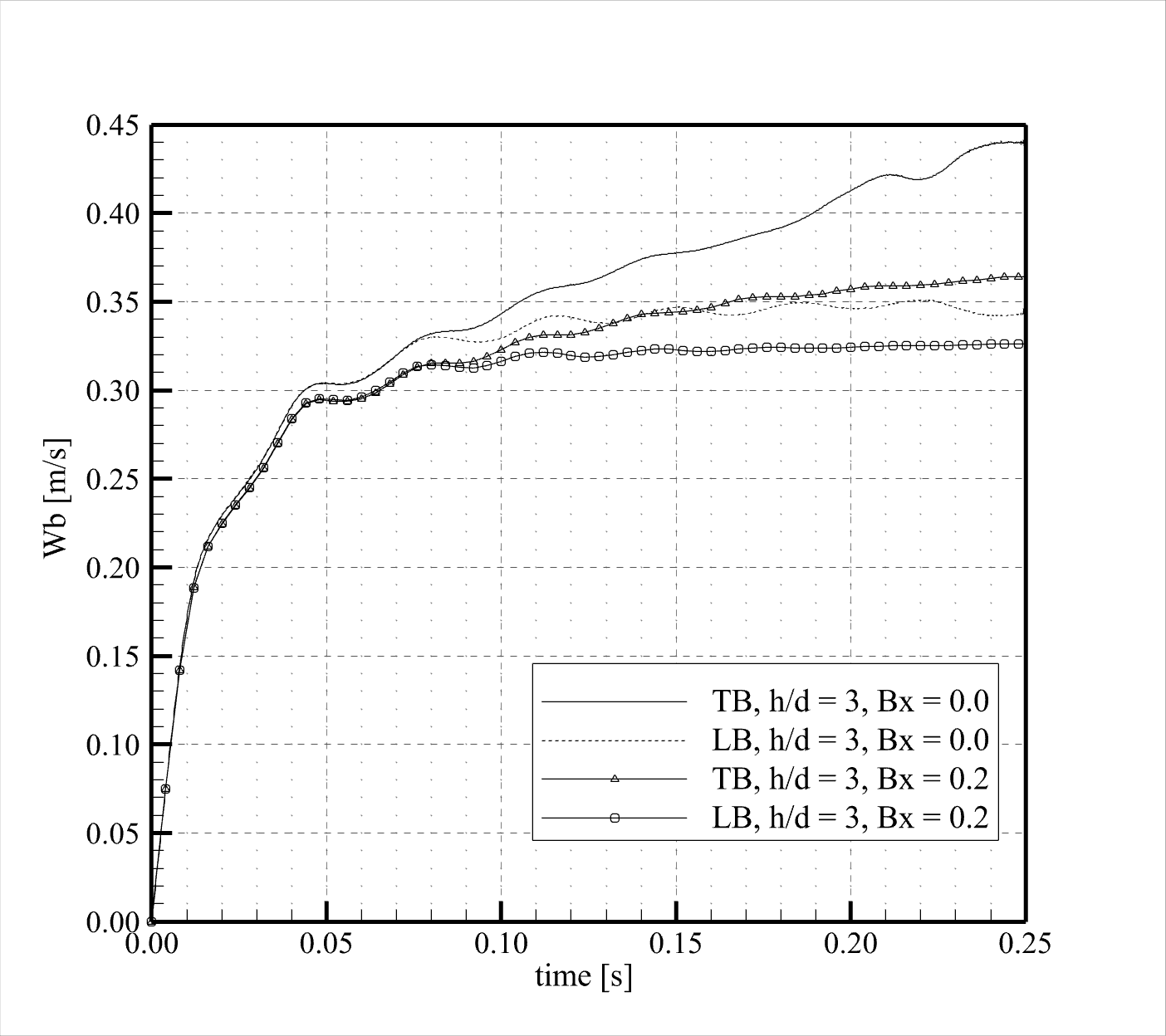}
		\caption{Rise velocity of leading and trailing bubbles for two magnetic fields}
		\label{fig:rise_vel_d_7mm_hd_3}
	\end{center}
\end{figure}

Figure \ref{fig:rise_vel_d_7mm_hd_3} shows the rise velocities of leading and trailing bubbles. Starting from stationary, rise velocity increases with time to go through fluctuations either to reach a steady state or continue to increase with time. The behavior of leading and trailing bubbles' rise velocities are different and also depends on the applied magnetic field. In the absence of magnetic field, rise velocities of both leading and trailing bubbles are identical for up to $t \approx 75 ~ms$. We can notice that the rise velocity of the leading bubble goes through rapid oscillations with the magnitude of oscillation is less than $0.01 m/s$. It should also be noted that the mean velocity of leading bubble is almost unchanged after $t \approx 75~ms$. On the other hand, the trailing bubble continues to accelerate for the total time considered in this study. Additionally, the trailing bubble's rise velocity is always higher than that of the leading bubble. 

In the presence of a magnetic field, rise velocities of bubbles are identical for up to $t \approx 80 ~ms$, and afterward, the curves diverge. Velocity of the leading bubble after $80 ~ms$ increases very slowly compared to before $t = 80 ~ms$ and reaches almost a steady-state. On the other hand, trailing bubble continues to accelerate, although with smaller rate than initial acceleration, to achieve a higher velocity than the leading bubble. Comparing these curves, it is evident that the rise velocity and their oscillations in the presence of a magnetic field decreases. \par 

\begin{figure}[H]
	\begin{center}
		\begin{subfigure}[b]{0.22\textwidth}
			\includegraphics[width=1\textwidth,trim=4 4 4 4,clip]{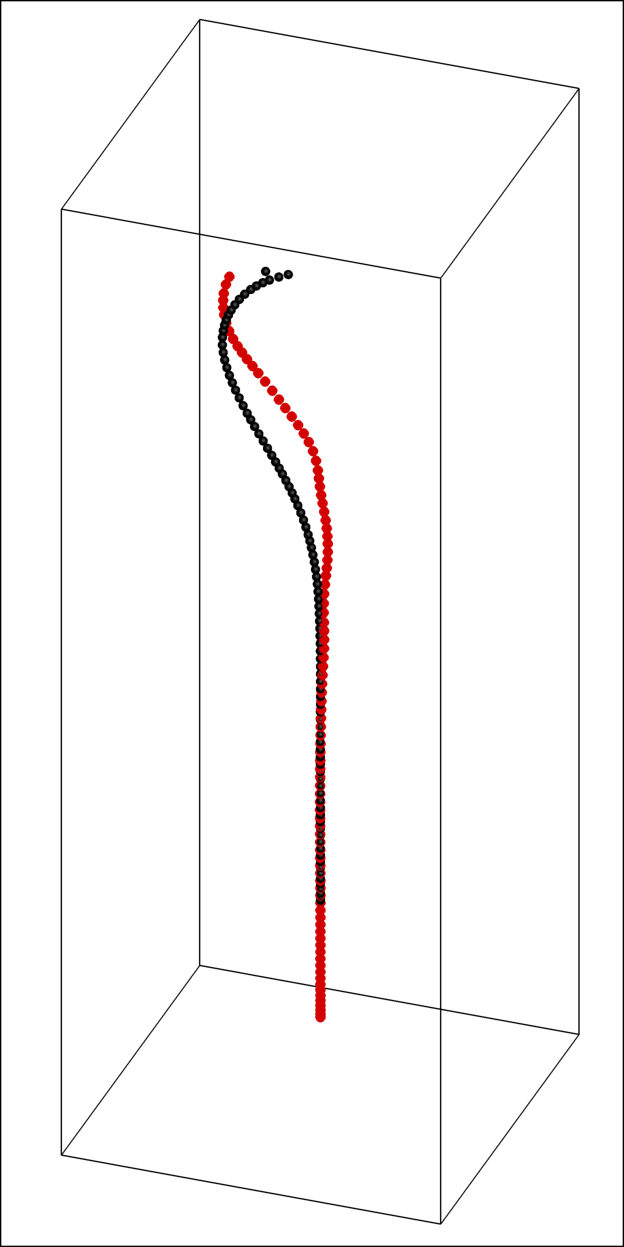}
			\caption{$\textbf{B}$ = 0}
			\label{fig:bubble_path_d_7mm_hd_3_B_0}
		\end{subfigure} %
		\begin{subfigure}[b]{0.22\textwidth}
			\includegraphics[width=1\textwidth,trim=4 4 4 4,clip]{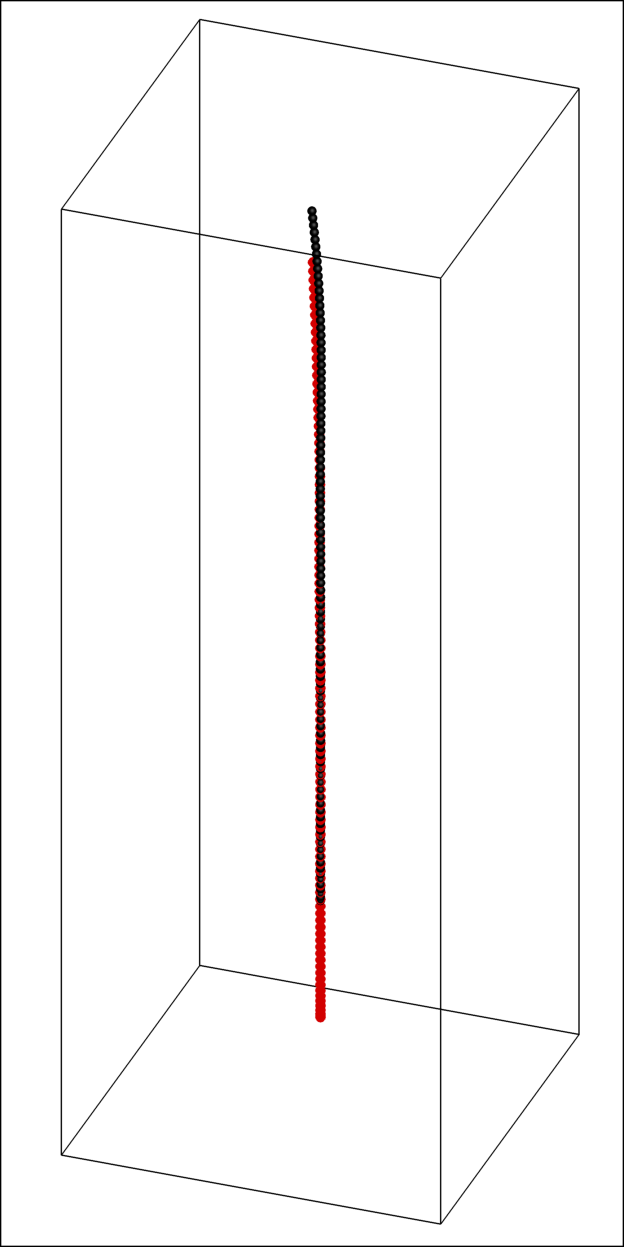}
			\caption{$\textbf{B}$ = 0.2}
			\label{fig:bubble_path_d_7mm_hd_3_B_0p2}
		\end{subfigure} %
		\caption{Rise paths of leading and trailing bubbles for two magnetic fields}
		\label{fig:bubble_path_d_7mm_hd_3d}
	\end{center}
\end{figure}

Figure \ref{fig:bubble_path_d_7mm_hd_3d} shows the trajectory of leading and trailing bubbles as they ascend -- red and black curves represent trajectories of trailing and leading bubbles, respectively. It can be noticed that in the absence of magnetic field both bubbles initially follow rectilinear paths turning into zigzags. Further, the trailing bubble deviates from rectilinear path earlier than the leading bubble. At the outset of ascend, trailing bubble has a smaller velocity and is away from the flow field modified by the leading bubble, therefore trailing bubble's shape changes from the spherical to asymmetric ellipsoid and does not exhibit surface oscillations. This is the primary reason for the initial rectilinear path of the bubble. As bubbles accelerate and reach a certain critical speed (or Reynold number), they start to shed symmetric vortices which later grows into asymmetric. This leads to a non-uniform pressure distribution behind the bubble, which pushes bubbles away from liquid column's center and migrates into a zigzag trajectory. 

When the magnetic field is included, both bubbles rise in the rectilinear paths for a longer time; however, towards the end of the simulation, they start to deviate and could potentially migrate into zigzag paths. This is in line with our observations from Fig. \ref{fig:iso_surfaces_d_7mm_hd_3_front}, the bubble reached a steady shape without any shape or orientation oscillations. To further understand the zigzag rise path and oscillatory rise velocity, we analyze the flow structures behind the bubbles. \par

\begin{figure}[H]
	\begin{center}
		\begin{subfigure}[b]{0.25\textwidth}
			\includegraphics[width=0.8\textwidth,trim=0 0 0 0,clip]{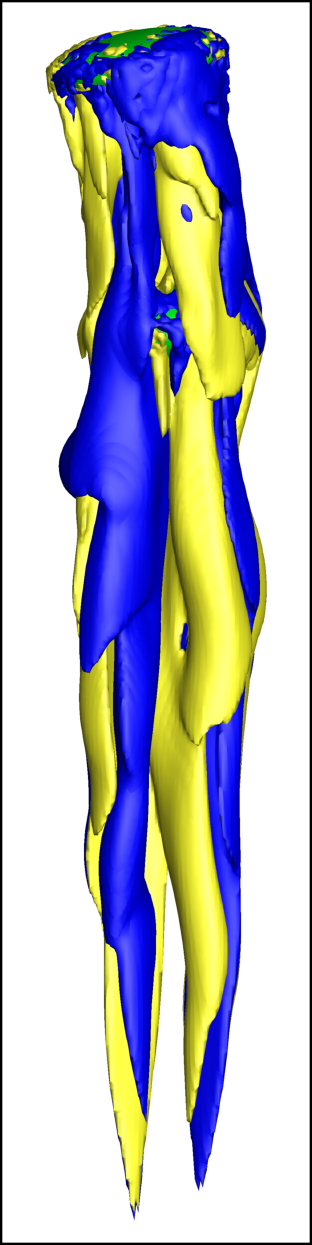}
			\caption{$\textbf{B}$ = 0}
			\label{fig:vorticity_d_7_hd_3_B_0p0}
		\end{subfigure} %
		\begin{subfigure}[b]{0.25\textwidth}
			\includegraphics[width=0.8\textwidth,trim=0 0 0 0,clip]{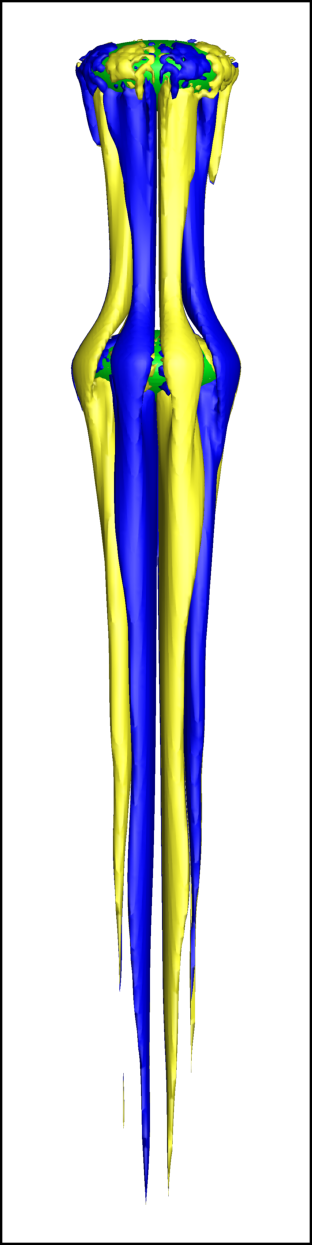}
			\caption{$\textbf{B}$ = 0.2}
			\label{fig:vorticity_d_7_hd_3_B_0p2}
		\end{subfigure} %
		\caption{Iso-surfaces of $\omega_z = \pm 25 $ at $t = 240~ms$ for two magnetic fields}
		\label{fig:vorticity_d_7mm_hd_3d}
	\end{center}
\end{figure}

Figure \ref{fig:vorticity_d_7mm_hd_3d} shows the front view of the iso-surfaces of $\omega_z$ at $t = 240~ms$ (The blue and yellow represents iso-surfaces of $\omega_z = 25$ and $\omega_z = -25$ respectively).  We can notice long elongated wake structures behind the bubbles with ``tail-like'' shape.  In absence of the magnetic field, these structures are flatter and entangled which shows that these counter-rotating vortices are interacting with each other and can disintegrate into smaller vortical structures. We can further notice non-uniform vortex shed from the leading bubble -- potentially modifying the flow-field observed by the trailing bubble.  This explains why rise path of trailing bubble deviates from rectilinear trajectory earlier than the leading bubble. In presence of the magnetic field, we notice symmetrical and more organized vortex shedding from the leading bubble. These structures interact and surround the trailing bubble, however, symmetry of the flow structures remain intact. The vortex shedding from both leading and trailing bubbles applies lateral forces on them and can push them away from the duct centerline which can result into zigzag trajectories. We can relate figures \ref{fig:iso_surfaces_d_7mm_hd_3_front}, \ref{fig:bubble_path_d_7mm_hd_3d} and \ref{fig:vorticity_d_7mm_hd_3d} and infer that the non-rectilinear motion of bubbles is related to the structure of the wake formed behind the bubbles. These instabilities in the wake cause an asymmetrical flow behind the bubble and lead to a zigzag motion.  Now we look at why vortical structures are more streamlined in presence of the magnetic field. \par
\begin{figure}[H]
	\begin{center}
		\begin{subfigure}[b]{0.48\textwidth}
			\includegraphics[width=0.98\textwidth,trim=4 4 4 4,clip]{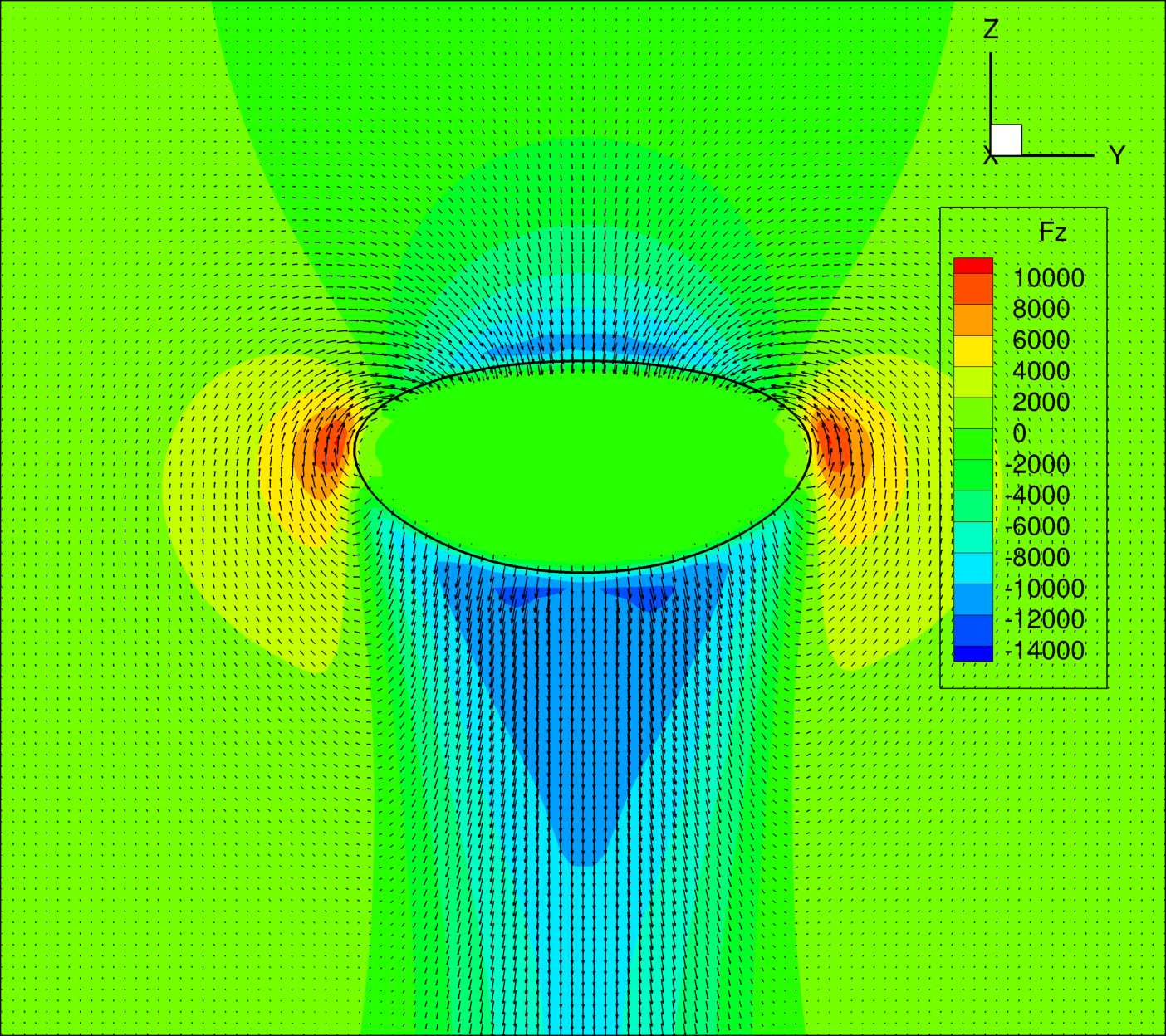}
			\caption{Vertical Lorentz force around leading bubble}
			\label{fig:LF_vec_LB_d_7mm_hd_3_B_0p2}
		\end{subfigure} %
		\begin{subfigure}[b]{0.48\textwidth}
			\includegraphics[width=0.98\textwidth,trim=4 4 4 4,clip]{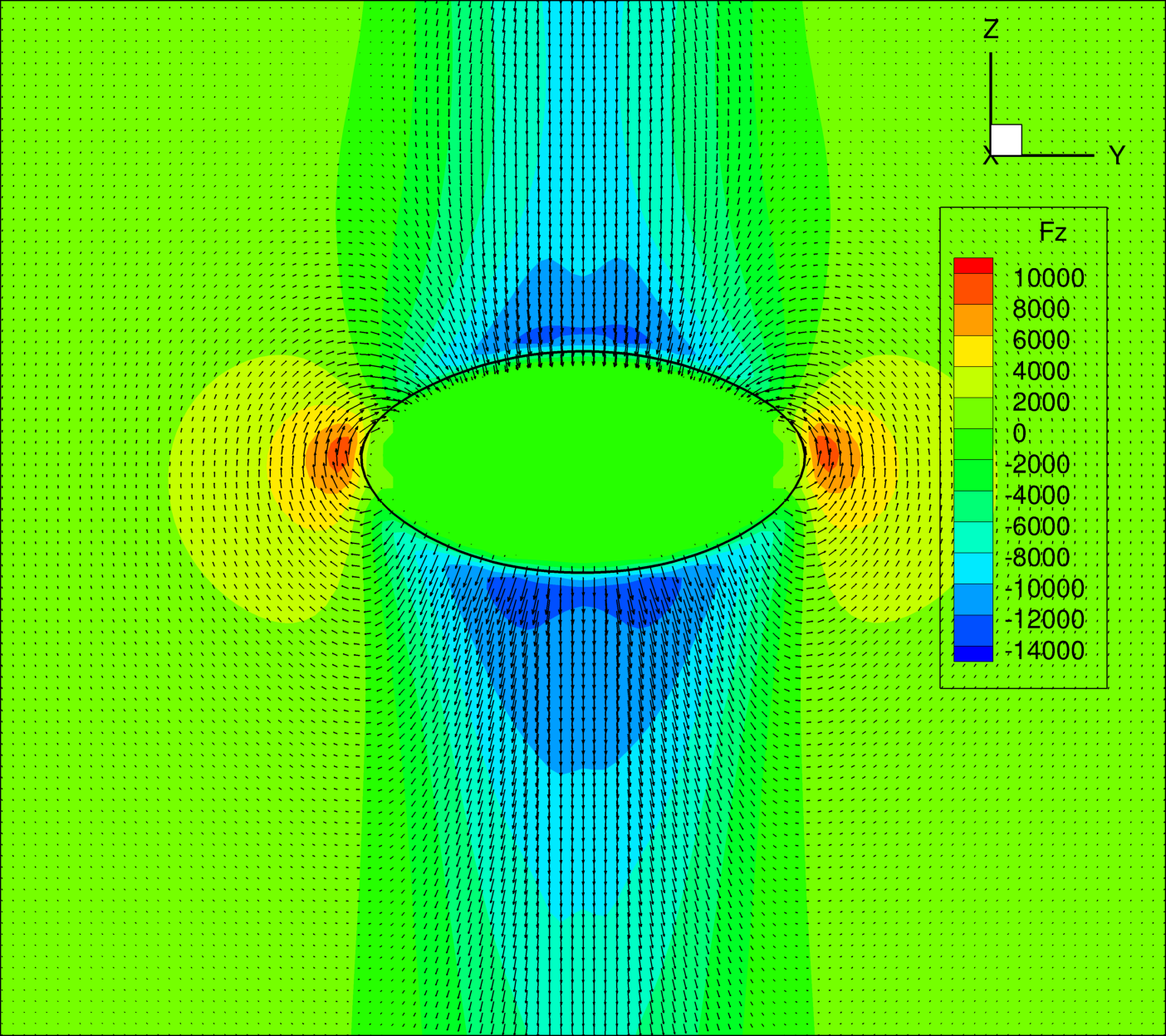}
			\caption{Vertical Lorentz force around trailing bubble}
			\label{fig:LF_vec_TB_d_7mm_hd_3_B_0p2}
		\end{subfigure} %
		\caption{Vertical Lorentz force distribution around bubbles on $yz-$plane at $t = 160~ms$ for $\textbf{B} = 0.2~T$}
		\label{fig:LF_d_7mm_hd_3d}
	\end{center}
\end{figure}
%

Figures \ref{fig:LF_vec_LB_d_7mm_hd_3_B_0p2} and \ref{fig:LF_vec_TB_d_7mm_hd_3_B_0p2} show the vertical Lorentz force $\textbf{F}_{lz}$ distribution overlaid with the Lorentz force vectors around leading and trailing bubbles on the $yz-$plane for a representative time of $t=160~ms$. From these two figures, we can see that the Lorentz force acts in the negative $z-$direction on both bubbles. However, the magnitude of the force is higher on the trailing bubble than that on the leading bubble. This will act to decelerate the trailing bubble more than the leading bubble. We can also see that the Lorentz force is symmetric to the vertical centerline. Therefore, the magnetic force will act to reduce the bubble wobbling. In Fig. \ref{fig:rise_vel_d_7mm_hd_3}, we observed that the trailing bubble velocity is higher than that of the leading bubble, which indicates that the effects of flow modification by the leading bubble on drag force is higher than Lorentz force.

\section{Conclusions}
\label{sec:conclusions}
We have studied the three-dimensional dynamics of two inline Argon bubbles rising in molten steel in the presence of a transverse magnetic field. We have used a VOF interface tracking method with the sharp surface force treatment to capture the interface and inclusion of surface tension forces accurately. The electromagnetic equations have been solved to capture the effects of the magnetic field. Then, we carried out two simulations of different magnetic fields for $7~mm$ bubbles placed $3d$ center-to-center distance apart. We analyzed the bubble deformation, rise path, rise velocity, and flow structures behind the bubbles. \par
We observe that the magnetic field strongly influences bubble deformations. For $\textbf{B} = 0.0$, both leading and trailing bubbles deformed into flatter ellipsoid with asymmetric top and bottom surfaces and went through shape oscillations. For $\textbf{B} = 0.2$, the leading bubble reaches a steady-state shape of an ellipsoidal bubble, whereas the trailing bubble goes through mild shape oscillations. Further, we observe that for $\textbf{B} = 0.0$, the bubbles initially rose in a rectilinear path but migrated into a zigzag path, and the amplitude of the path is nearly twice the bubble diameter. For $\textbf{B} = 0.2$, bubbles rose in a linear path for most of their ascent and migrated away from the center towards the end.

The rise velocity goes through oscillations in both magnetic field cases; however, in the absence of the magnetic field, the magnitude of the velocity is significantly larger. We found that these oscillations are related to the shape oscillation of the bubble and a zigzag rise path. In both cases, leading and trailing bubbles initially ascended with identical speeds. Once maximum deformation was achieved, the trailing bubble rose faster than the leading bubble. 

%
%

\bibliographystyle{unsrtnat}	
\bibliography{ms.bib}

\end{document}